ECOLE NORMALE SUPERIEUR

Synthèse de l'activité scientifique en vue de l'obtention

d'une Habilitation à Diriger des Recherches

**Andrei ZINOVYEV**

U900 Institut Curie/INSERM/Mines ParisTech

"Bioinformatics and Computational Systems Biology of Cancer"

Institut Curie, rue d'Ulm 26, 75248, Paris

---

# *DEALING WITH COMPLEXITY OF BIOLOGICAL SYSTEMS: FROM DATA TO MODELS*

---

*Soutenu le 04 Avril 2014*

*Devant le jury compose de :*

*Président: Prof., Dr. Denis Thieffry*

*Rapporteurs:*

*Prof., Dr., Lodewyk Wessels*

*Dr. Marie-France Sagot*

*Dr. Dirk Drasdo*

*Examinateurs :*

*Prof., Dr. Alexander Gorban*

*Prof., Dr. Ovidiu Radulescu*

*Prof., Dr. Daniel Gautheret*

*Dr. Emmanuel Barillot*

# *Contents*

# *ABSTRACT*

In this synthesis I give an informal review of the scientific projects in which I was involved during last 12 years. A common transversal topic of developing and applying theoretical methods for dealing with biological complexity is suggested for connecting these projects together.

Four chapters of the synthesis represent four major areas of my research interests: 1) data analysis in molecular biology, 2) mathematical modeling of biological networks, 3) genome evolution, and 4) cancer systems biology.

The first chapter is devoted to reviewing my work in developing non-linear methods of dimension reduction (methods of elastic maps and principal trees) which extends the classical method of principal components (PCA). Also I present application of advanced matrix factorization techniques based on blind source separation to analysis of cancer data.

The second chapter is devoted to the complexity of mathematical models in molecular biology. I describe the basic ideas of asymptotology of chemical reaction networks aiming at dissecting and simplifying complex chemical kinetics models. Two applications of this approach are presented: to modeling NFkB and apoptosis pathways, and to modeling mechanisms of miRNA action on protein translation.

The third chapter briefly describes my investigations of the genome structure in different organisms (from microbes and model organisms to human cancer genomes). Unsupervised data analysis approaches are used to investigate the patterns in genomic sequences shaped by genome evolution and influenced by the basic properties of the environment.

The fourth chapter summarizes my experience in studying cancer by computational methods (through combining integrative data analysis and mathematical modeling approaches). In particular, I describe the on-going research projects such as mathematical modeling of cell fate decisions and synthetic lethal interactions in DNA repair network.

The synthesis is concluded by listing major challenges in computational systems biology, connected to the topics of this text, i.e. dealing with complexity of biological systems.

# *RESUME*

**APPREHENDER LA COMPLEXITE DES SYSTEMES BIOLOGIQUES : DES DONNEES AUX MODELES**

Dans cette synthèse je présente une revue informelle des projets scientifiques auxquels j'ai participé pendant ces 12 dernières années. Le développement et l'application de méthodes théoriques pour appréhender la complexité biologique sont proposés comme sujet transversal connectant ces projets.

Les auatre chapitres de cette synthèse représente quatre domaines majeurs de mon intérêt scientifique: 1) l'analyse de données en biologie moléculaire ; 2) la modélisation mathématique de réseaux biologiques ; 3) l'évolution des génomes ; 4) la biologie des systèmes du cancer.

Le premier chapitre présente mon travail sur le développement des méthodes non-linéaires de réduction de la dimension des données (méthodes des cartes élastiques et les arbres principaux). Ces méthodes généralisent l'approche classique d'analyse en composantes principales. Ensuite, je décris l'application des méthodes avancées de factorisation de matrices basées sur la séparation aveugle des sources.

Le deuxième chapitre présente une approche pour analyser la complexité de modèles mathématiques en biologie moléculaire. Je décris l'idée principale de l'asymptotologie des réseaux de réactions chimiques qui vise à la dissection et la simplification de modèles complexes de la cinétique chimique. Deux applications de cette approche sont présentées : la modélisation des voies de signalisation NFkB et de l'apoptose et la modélisation des mécanismes d'action des microARN sur la traduction de protéines.

Le troisième chapitre introduit brièvement mes investigations sur la structure du génome de différents organismes (des microbes et des organismes models aux génomes de cancers humains). L'analyse non-supervisée des données est appliquée pour étudier les patterns dans les séquences génomiques qui sont affectées par l'évolution génomique et par les propriétés de l'environnement.

Le quatrième chapitre résume mon expérience dans les études du cancer par des approches computationnelles (par la combinaison d'analyses intégratives des données et de modélisation mathématique). En particulier, je décris des projets en cours comme la modélisation mathématique du destin cellulaire et les interactions synthétiques létales dans le réseau de réparation d'ADN.

La synthèse est conclue pas la formulation des défis majeurs en biologie des systèmes qui sont reliés au sujet de ce texte, la lutte contre la complexité des systèmes biologiques.

## *Preface*

*To my parents and*
*to my fantastic wife*

In this short text I summarized the main scientific projects in which I was involved for the last twelve years, since I defended my PhD thesis in computer science. This was not a simple task for me, because it was difficult to find a common transversal theme through very different scientific problems and questions. After some reflection, I realized that this can be the concept of *fighting with complexity of biological systems*. During years, I was collecting an arsenal of methods from data mining and theory of dynamical systems for this mission but I also seriously studied the biology itself. Some of the methods were developed with my participation, some I borrowed but adapted to the type of biological problems I have studied.

My PhD thesis was devoted to developing a statistical methodology: methods for non-linear dimension reduction and data approximation. I was applying it in many different domains, including the bioinformatics task of de novo gene positions finding in genomic sequences. This experience taught me several important lessons: (1) *Often the problem can be solved by a much simpler (compared to the state of the art) method* with the same if not better efficacy; (2) *Meaningful application of mathematics requires rather serious understanding of the subject field* (i.e., biology); (3) *Solving the problem and method development should go hand in hand*, because there are no universal methods. I followed these principles since then, and I find this way of doing interdisciplinary science practical and useful. Metaphorically speaking about the subject of this synthesis, *to fight with complexity you have to know the enemy and you have to know the weapon*.

This synthesis is divided into four chapters, roughly representing my current research interests: data mining, mathematical modeling of biological systems, understanding genome evolution and computational cancer systems biology. As I have explained, it was not possible for me to separate “methodological” and “scientific” aspects: they are indeed glued together, so I did not introduce any artificial splitting between them, though first two chapters are devoted more to my methodological contributions.

# *INTRODUCTION*

We need to fight with complexity when we study biology because biological systems “resist” to our attempt to understand, hence, efficiently manipulate them. But what does it exactly mean - “fighting with complexity”?

Many different things stay behind the notion of complexity (here and further I deliberately avoid giving a precise definition of this hotly discussed term): the number of elements or connections between them, number of intrinsic degrees of freedom, non-triviality of behavior, non-linearity of mathematical equations, difficulties with abstraction, etc. Some researchers associate the notion of complexity with *non-linearity or large dimension*. Others connect complexity with *emergence and self-organisation* (Miller and Page, 2007). Some point out that the main challenge on this way is to distinguish *complicated* and *truly complex systems*, though no consensus view on the nature of this distinction exists in the community. Complexity of biological systems is tightly connected to their *robustness* and the history of their *evolution* (Wagner, 2005; Barillot et al, 2012).

In his “millennium” interview, Stephen Hawking said “I think the next century will be the century of complexity”. What is meant here is that in the XXI century most of scientific effort will be devoted not to discovery of new scientific laws; but rather, knowing the fundamental laws, to understanding how complex systems are assembled and function. Citing another paper on complexity theory, “a new scheme of actions became dominant in the struggle with complexity. The *complexity is recognized as the gap between the laws and the phenomena*. We assume that the laws are true. We can imagine a ‘detailed’ model for a phenomenon but because of complexity, we cannot work with this detailed model. We can imagine a detailed kinetic equation for a reaction network but cannot find reaction rate constants and cannot work with this large system even if it is true” (Gorban and Yablonsky, 2013).

In this “anthropic” view, complexity is presented as an obstacle for a human mind, equipped with modern technology, to interpret behavior of complex systems based on a set of simple laws, using logical/mathematical/computational deductive reasoning.

We can look into the most general classification of methods that were developed for fighting with complexity. Very roughly, there are two classes of these methods: based on *model/dimension reduction* and based on *self-averaging* (Gorban, 2009; Barillot et al, 2012). To understand this classification, we can think of a complex phenomenon as an object existing in a multi-dimensional space (e.g., a set of points or trajectories or a vector field). Our perception of this object is inevitably low-dimensional because our mind is organized by representation of our motion in three-dimensional space and the convenient static visualisation is two-dimensional. Therefore, we can represent our perception as a projection of the object from high-dimensional to low-dimensional space. A biological function can be also considered as a projection of its high-dimensional microscopic detailed description onto a low-dimensional space where it is manifested at macroscopic level. Let us try to imagine what one can observe through such a projection.

*The reducible complexity* model states that despite the fact that the complex object is embedded in a high-dimensional space, intrinsically, it remains low-dimensional with a relatively small number of degrees of freedom. Reducible complexity of data often means existence of lower-dimensional *principal manifolds*. Reducible complexity of dynamical systems is manifested in low-dimensional intrinsic structure of their attractors or existence of low-dimensional *invariant manifolds*. Another frequent type of reducible complexity is a system's structure following some relatively simple organisational principle. One of the most common principles is the *hierarchical organisation*. In biological networks, this type of hierarchical reducible complexity is revealed in the existence of modules, compartmentalisation, multiple concentration and time scales (Radulescu et al, 2006). In physiology, it can be seen as the construction of an organism from organs, tissues and cells.

By contrast, *self-averaging complexity* is associated with truly high-dimensional objects that do not possess any intrinsic low-dimensional simple structure. However, after projection on most of the low-dimensional screens, the self-averaging object will look very similar. A good mathematical metaphor for this type of complexity is a multi-dimensional shape (a hypercube, a hypersphere) uniformly sampled by points. After projection on any two-dimensional plane, most of the points will be located very close to the center of the projection distribution. Moreover, the distribution of the projected points will be very close to a normal one (i.e. Gaussian). In statistical physics, this corresponds to the well-known Maxwellian distribution. Generally speaking, this is an example of Gromov's measure of concentration phenomenon: truly high-dimensional objects look very small (concentrated) after projection onto a low-dimensional space and most of the distributions become almost normal after projections on the low-dimensional subspaces (Gromov, 1999).

The systems whose behavior cannot be reduced to a relatively simple view or do not self-average, possess “wild” complexity, with which we do not have scientific tools to fight with. We can only observe some of their features and try to reproduce them by engineering more or less complex *models* of their behavior. In the field of systems biology, complex datasets (e.g., genome-wide measurements, omics data) and complex models (e.g., global interaction networks) started to appear at increasing rate. An important question is to what extent we are able to simplify them in order to understand and get control over biological systems behavior. Below I present different projects in which this question was posed in some way, and specific computational or mathematical tools were used to answer it.

# 1. THE BIG DATA OF MOLECULAR BIOLOGY

## 1.1 Curse of data quantity and of data dimensionality

The current state of molecular biology and genetics is characterized by unprecedented influx of quantitative data. Twenty years ago, all the data that bioinformaticians possessed was relatively small amount of fragmentary genetic sequences for which they applied carefull and intellectually challenging analysis (such as estimation of the mutation rates). Modern biotechnologies allow generating the data at the exponential rate which exponent is larger than the exponent in the Moore's law describing the growth of the computational power in hardware (Barillot et al, 2012). This inevitably leads to a situation when storing and analyzing the data becomes more expensive than producing them.

The technological challenge of efficient storage, compressing and pre-treating the data is accompanied by a scientific challenge of using the data in order to extract a useful knowledge from them. A biological sample can be characterized by increasing number of quantitative features. In the beginning of microarray era one had few tens of thousands of measurements per sample. Using modern sequencing techniques, the number of extractable numerical features such as the number of RNA counts, different forms of RNA, mutations or epigenetic modifications of DNA, variations in DNA sequence, grew up by several orders of magnitude. This leads to a methodological problem called "*small n, big p*" problem (where *n* is the number of samples and *p* is the numbers of quantitative features describing the sample) which can be illustrated in the following abstract way.

Imagine that a biological sample is a point in a multidimensional space whose dimension equals to the number of measured numerical features. A collection of samples is represented as a cloud (discrete distribution) of points in this space. If this distribution is truly multidimensional (there is no a low-dimensional subspace around which this cloud is concentrated) then *the average distance between a point and its closest neighbor* and *the average distance between all pairs of points* are comparable. Imagine a ranking of points from a selected point sorted by distance. In the truly multidimensional situation, adding a small amount of noise to the position of data points risk to change this ranking drastically. This is the essence of the *curse of dimensionality.* Most of the statistical methods, including the simplest *k*-nearest neighbors classifiers and linear predictors, performs very poorly in such conditions, their parameters are too sensitive to imprecisions in data measurements or to random removal of a small fraction of samples from the dataset.

One of the manifestations of the curse of dimensionality is difficulty of finding recurrent patterns in the data such as frequently repeated events. In cancer, for example, most of the mutations which can in principle serve as biomarkers of, for example, the treatment success, are present in small numbers which hampers evaluating their statistical significance and requires an enormous (and not always feasible) amount of biological samples.

Curse of dimensionality is one of the sides of the biological complexity. Typical methods which were developed to fight with this type of complexity are feature selection methods, regularization and dimension reduction. During the last twelve years, I have made several contributions in this direction, developing methods of non-linear dimension reduction, factor analysis, and providing recipes for measuring the data complexity.

## *1.2 Linear and non-linear data dimension reduction*

For data approximation, the notion of the mean point can be generalized by more complex types of objects. In 1901 Pearson proposed to approximate multivariate distributions by lines and planes (Pearson, 1901). In this way the Principal Component Analysis (PCA) was invented, and nowadays it is a basic statistical tool. Principal lines and planes go through the "middle" of multivariate data distribution and correspond to the first few modes of the multivariate Gaussian distribution approximating the data.

I worked on extending the data approximation methods for other types of principal objects: first of all, by non-linear manifolds of various topologies. This lead to the creation of the method of elastic maps for constructing principal manifolds (Zinovyev, 2000; Gorban et al, 2000,2001; Gorban and Zinovyev 2005; Gorban et al, 2008A). This method is implemented in several software packages available online (http://bioinfo-out.curie.fr/projects/vidaexpert, http://bioinfo-out.curie.fr/projects/elmap, http://bioinfo-out.curie.fr/projects/vimida ).

Further extension of this approach led to the idea of approximating datasets by arbitrary graphs (principal graphs) (Gorban and Zinovyev, 2005, 2007; Gorban et al, 2007, 2008B; Gorban and Zinovyev, 2009, 2010). It was suggested to define such graphs by applying topological grammars. The simplest possible grammar leads to the *method of principal trees*.

In 2006, I participated in the organization of an international workshop "*Principal manifolds for data cartography and dimension reduction*". The following scientific challenge was suggested to the participants of the workshop: "*… to have a look at three datasets containing results of a high-throughput experimental technology application in molecular biology (microarray data) … any kind of message about the structure of the point distribution and relation of this structure to the proposed ab initio gene and sample classifications is interesting.*" (http://www.ihes.fr/~zinovyev/princmanif2006/). This task led to an interesting discussion at the workshop and publishing a book on principal manifolds (Gorban et al, 2008A), which became a standard reference in the field.

Recently, we used approximation of datasets by principal trees in order to evaluate the complexity of the datasets (Zinovyev and Mirkes, 2013). We introduced three natural types of data complexity: 1) geometric (deviation of the data's approximator from some ``idealized" configuration, such as deviation from *harmonicity*); 2) structural (how many elements of a principal graph are needed to approximate the data), and 3) construction complexity (how many applications of elementary graph transformations are needed to construct the principal object starting from the simplest one). We computed these measures for several simulated and real-life data distributions and showed them using the “accuracy-complexity” plots, helping to optimize the accuracy/complexity ratio.

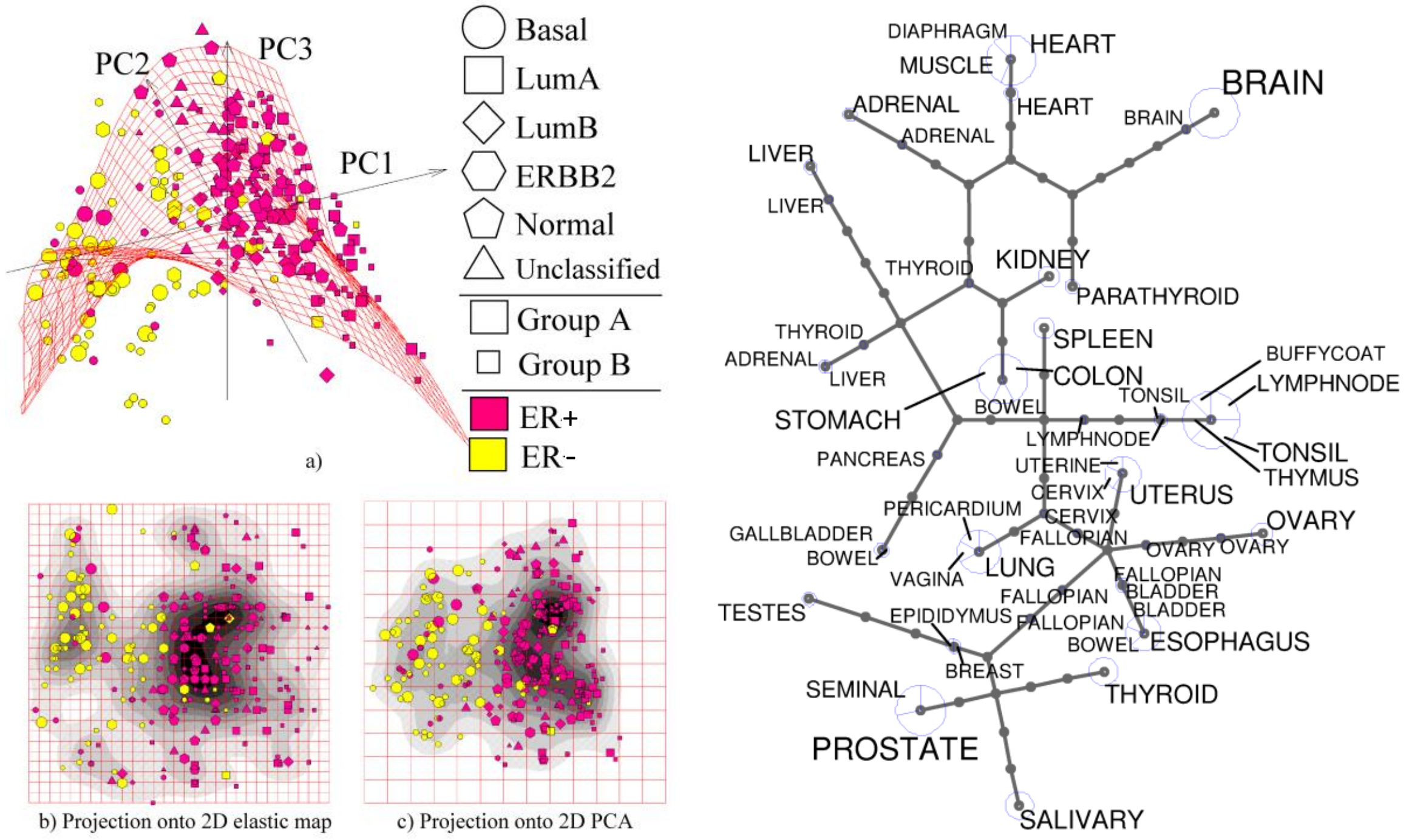

Figure 1. Linear, non-linear and graph-based approximations of biological datasets (microarray data, transcriptomes). Right: comparison of linear and non-linear principal manifolds for visualization of the structure of distances between breast tumor samples (from Gorban and Zinovyev, 2008). Left: visualization of a dataset of normal tissue samples (from about 100 tissues), the structure of the distances is represented by a tree structure (from Gorban et al, 2008A).

Reduction of the data dimension represents the first step towards fighting with complexity. Reducing the number of variables by eliminating uninformative ones or lumping the variables into weighted combinations helps in extracting the knowledge from the data.

## *1.3 Blind source separation of biological signals*

“Small n, big p” problem can be transposed. One can be interested in studying the numerical features (such as expression of a particular gene) across a number of biological samples. This situation can be represented as a distribution of $p$ points in $n$-dimensional space, where $p \gg n$. This analysis answers the following question: *which biological or technical factors affect the genome-wide measurements* (for example, the transcriptome)? The biological factors can be activities of transcription factors or other various influences coming from a particular intercellular context or from the environment. The technical factors can be various biases connected to the preparation or even extraction of the samples, or to the technology used. The combination of factor activities regulates gene expression in a complex (and unknown) function. As the first approximation, we can assume that this function is linear:

$$\text{Expression(gene } i\text{, sample } s) = \Sigma_{i=1..m}\, a_{Fj}{}^{gene\ i}\,\text{Activity}_{Fj}(\text{sample } s).$$

Such a linear decomposition of the gene expression table depends on the conditions which are put onto the gene responsiveness coefficients $a_{Fj}{}^{gene\ i}$ connected to a particular signal $F_j$ and the activities $\text{Activity}_{Fj}$ of this signal in a sample *s*. Orthogonality or uncorrelatedness condition leads to the standard Singular Value Decomposition problem. Requirement of statistical independence of the signals leads to the Independent Component Analysis (ICA) method, which was initially developed for signal processing field (see review on application of this method for cancer research in (Zinovyev et al, 2013C)).

We applied this approach to comprehensively characterize the activity of biological factors in transcriptome of bladder cancer, and compared them to the signals that can be detected in other cancer types (Figure 2).

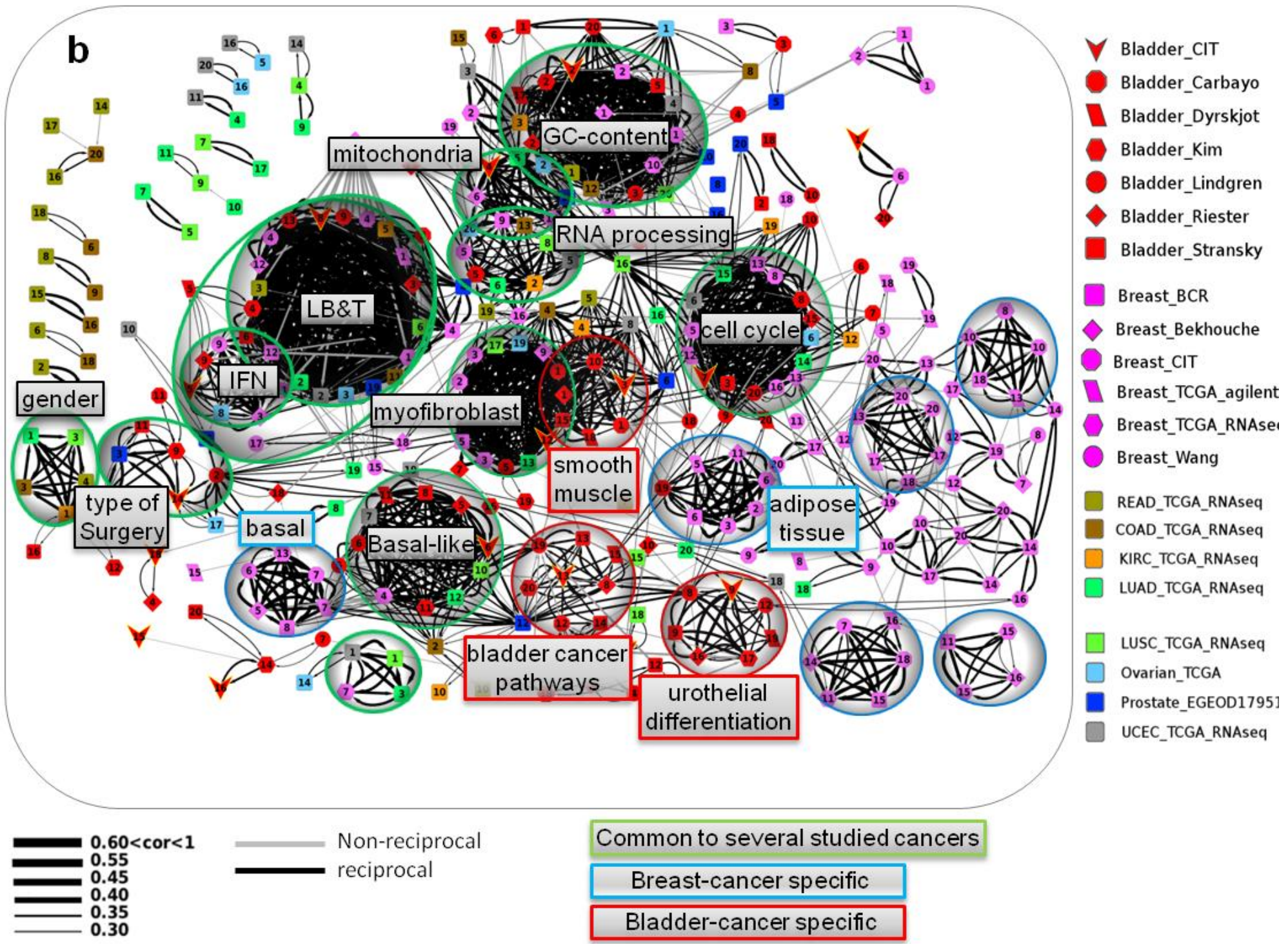

Figure 2 Comprehensive characterization of transcriptomes of collections of tumour samples in different solid cancers (from Biton et al, 2013). The image is a correlation graph representing the correlations between independent factors identified in various datasets (node shapes) and cancer types (node colors). The text annotations denote several clusters of components (cliques) representing reproducible biological signals which origin is relatively well-understood. For example, LB&T represents a set of components connected to the reaction of immune system (presence of lymphocytes, B- and T-cells in the tumour sample). The “Type of surgery” clique represents components induced by different surgical protocols for extracting the tumour, etc.

We interpreted the inferred independent components by studying their association with predefined gene sets, and also with the clinical, pathological and experimental data, mutations, and copy number alterations available for the these tumors. We also applied ICA to publicly available gene expression datasets, composed of six series of bladder cancers and 13 series of cancers of other types. Our results showed that ICA allows: the identification of gene expression data which are influenced by technical biases; the identification of processes that were either cancer type-specific or common to several types of cancer, the characterization and the comparison of predefined tumor subgroups and the identification of genes involved in carcinogenesis. One of our findings, the potential pro-tumorigenic role of the nuclear receptor PPARG (peroxisome proliferator-activated receptor gamma) in bladder tumorigenesis, was confirmed by functional studies (Biton et al, 2013).

Meta-analysis of cancer transcriptomes with use of blind source separation methods, allows answering important questions of what are the most reproducible factors affecting cancer transcriptome, among various types of cancers? For example, it can be demonstrated that the factor associated with organization of tumor microenvironment (extracellular matrix properties, expression of metalloproteases, various aspects of cell-cell and cell-matrix adhesions) is the most reproducible among breast cancer transcriptomic datasets. At the same time, in a set of four solid carcinomas (breast, ovary, lung and prostate cancer), the most reproducible common signal is associated with immune response and infiltration of T-cell in the tumoral tissue.

# 2. MATHEMATICAL MODELS OF BIOLOGICAL SYSTEMS: COMPLEX OR COMPLICATED?

## 2.1 Asymptotology of reaction networks

Biological systems show complex behavior, where complexity is manifested in the difficulty to predict what will happen with a particular biological system or its element as a result of a perturbation. The biological systems are constructed from microscopic parts which are not directly observable. Hence, any observation on their behavior represents a complex function integrating many molecular components and their interactions. Therefore, it is inevitable in molecular biology to introduce imaginary models connecting the microscopic and macroscopic behavior. The notion of a model is well established in molecular biology. Mathematical approach provides a formal language in which such models can be formulated and analyzed. The model serves as a “theoretical microscope” allowing to look mentally at the unobservable molecular mechanisms from their macroscopic manifestations assuming the validness of the basic laws that govern the molecular interactions.

The most fundamental basis for constructing the formal biological models are the laws of (bio)physics and (bio)chemistry. One can assume that these laws are well-known. However, these laws by themselves do not explain functioning of biological systems. The laws of physics, as Galileo pointed out, tell you that “there are no flying elephants but they cannot help you to draw the path that brought walking elephants to Earth”. The main challenge for molecular biology is to connect the basic laws of nature to complex organization of biological organisms. This is another challenge of fighting with complexity of biological systems.

Most mathematical methods for modelling regulatory mechanisms are based on formal methods of chemical kinetics developed for studying chemical or biochemical systems. Historically, these methods were introduced to deal with relatively small and well-defined systems of chemical reactions. It is clear now that the biochemical cascades involved in cellular signalling can be characterised by large (a few hundreds of components) and extra-large (more than one thousand components) complex hierarchical structure and by multiscale temporal and spatial behaviour. Modelling of large biochemical networks, based on standard mathematical approaches, faces obstacles such as incompleteness of network description (structural and parametric), lack of exact knowledge of kinetic parameters, fuzziness of borders of the classical biochemical pathways, intensive pathway cross-talk, instabilities and poor scalability of numerical solvers. These peculiarities are not taken into account explicitly in the general mathematical methods applied today in most systems biology applications, which crucially limits the success of mathematical modelling in the field of systems biology of human diseases such as cancer.

Constructing large models of biochemical reactions gives serious advantages for realistic mathematical modelling of signalling pathways. Firstly, it affords the possibility of representing the complexity of molecular mechanisms without neglecting or a priori over-simplifying their components. This could be essential for understanding variability of individual response to treatment that cannot be captured by fixing a generic simple framework. Secondly, it allows the combination of separate existing mathematical models into comprehensive master models, when interaction between various biological mechanisms (for example, cell cycle and apoptosis) is known and cannot be neglected. Thirdly, it allows easier comparison between modeling and high-throughput data, since these data are available at the level of individual and elementary molecular entities. Lastly, determination of kinetic parameters can be less difficult for elementary molecular mechanisms as opposed to more abstract mechanism representations.

However, complex mathematical models can be as intractable as the biological systems themselves. With a very complex model, a researcher can observe possible system behaviour from numerical simulations, but he or she is not able to predict changes in the model dynamics as a response to changing its parameters values. Empirically, each particular numerical simulation can show very simple dynamical properties (for example, almost linear relaxation dynamics). In this case, this dynamics can be described by very few key parameters. The problem is in that the recipe for finding these parameters is usually not known.

Another problem in dealing with complex models consists in that usually they describe relatively simple dynamics of molecular entities described in a biological experiment. The parameters of complex models fitted to such experimental results will be highly undetermined (Hlavacek, 2009). The solution could be to reduce the complexity of a mathematical model to the level of complexity of experimental data, and fit only the necessary key parameters, which are usually some functions of the parameters in the detailed model.

All this dictates a need in developing methods of computational modeling allowing to work with large and incompletely characterized networks of biochemical interactions. In 2010, with my participation, a consortium of researchers was organized with aim to formulate this challenge explicitly and provide methods and software for dealing with large networks ( http://xxlnet.org/ ).

My contribution to this field consists in developing and applying analytical and computational methods allowing to simplify the equations of chemical kinetics in molecular biology. These methods can serve as a basis for the theory of *asymptotology of chemical reaction network*s (Gorban et al, 2010). Following Kruskal (1963), asymptotology is ‘‘the art of describingthe behavior of a specified solution (or family of solutions) of a system in a limiting case… The art of asymptotology lies partly in choosing fruitful limiting cases to examine… The scientific element in asymptotology resides in the non-arbitrariness of the asymptotic behavior and of its description, once the limiting case has been decided upon’’. Known methods of model reduction (Quasisteady-state, Quasiequilibrium asymptotics, lumping approaches, methods based on limiting reaction steps) are examples of finding simple asymptotic solutions of complex chemical kinetics equations, hence, they form the theoretical basis for asymptotology of reaction networks.

The interpretation of the asymptotology principles might be the following. Let us imagine a complex model of a large biochemical reaction network, characterized by a particular set of kinetic parameter values and some dynamics. Asymptotology claims that in many cases the dynamics of the model will not be as complex as it might be expected from the model size or its non-linearity. For example, in any given moment of time, most of the subsystems in the model will be characterized by the simplest quasi-linear relaxation, and only a minor subset of interactions will be described by truly non-linear equations. However, this subset might change in a different time point: a complex system “walks” through its simpler subsystems.

Asymptotic description of the dynamics consists in 1) for a given set of parameters, decomposing the complex dynamics into periods (epochs), each of which can be described in relatively simple terms (asymtptotic, based on neglecting some parameters or quantities); 2) for all admissible sets of paratemers, listing a set of possible asymptotic behaviors. Asymtotology provides tools for constructing this description for some classes of complex and large networks. Ideally, asymptotic solutions should be simple or trivial, and even analytically tractable. This allows understanding the system properties and predicting the possible changes of dynamical properties of the model as a response to a change of parameter values.

In biological terms, it means that any particular system response is characterized by a relatively simple sequence of events which are tractable and comprehensible. The complexity of the biological system arises from its ability to respond to many different types of signals and perturbations, thus, holding a capability of many simple but different dynamics.

Conceptually, it remains unclear if this view is applicable to the functioning of *real biological reaction networks*. If it is true then the *biological complexity can be dissected* as a superposition of relatively simple behaviors. In the opposite situation, the biological complexity is "wild", and not tractable. Imagine a well-defined large biochemical system, and a sufficient amount of experimental data to completely determine its parameters. The form of the distribution of these parameters will give a first hint to the system complexity: if they are distributed on the log scale then the complexity is most probably "dissectable". Secondly, one can investigate the properties of the vector fields describing the system dynamics. If it will be characterized by existence of low-dimensional attractive slow manifolds then the complexity is "dissectable". A method for making such a test was suggested in one of my works (Radulescu et al, 2007).

Our experience shows that most of the existing mathematical models (collected, for example, in BIOMODELS database, http://www.ebi.ac.uk/biomodels/ ) trained on some real experimental data are "dissectable" in terms of complexity; though a separate project is needed to demonstrate it formally. However, this can reflect only our way to fit the models to data, or to the limitations of experimental techniques. In principle, it is meaningful to ask the question if the natural selection leads to "wild" or "dissectable" complexity in biochemical reaction networks. Conceptually, this question is close to studying the evolution of network robustness (Radulescu et al, 2008, 2012; Barillot et al, 2012).

Asymptotology or model reduction methods allow fighting with biological complexity in several ways. Firstly, they allow making the model equations simpler and more tractable (analytically or numerically). Secondly, they allow determining the key model parameters and their relation to the parameters of the complete model. Thirdly, they allow dissecting model complexity into a set of simple models, and match each biological observation to a possible asymptotic model dynamics. Fourthly, they allow predicting how to switch between different asymptotic (qualitatively different) modes of behavior. Finally, they allow to decide if a mathematical model is truly ("wildly") complex of just complicated ("dissectable").

During last years, I participated in producing two theoretical results in this field: a *method of invariant grids* for computing approximations to slow manifolds (Gorban et al, 2004A; Gorban et al, 2004B; Gorban et al, 2005A), and a method of reducing a system of equations describing a *network of monomolecular reactions* in the case of well-separated kinetic constants (Radulescu et al, 2008). The details of the theoretical description of the methods are described in the corresponding publications. Interestingly, these methods have tight connection to quite abstract mathematics such as tropical algebras and "model tropicalization" (Radulescu et al, 2012) or the notion of dominant system in dynamical systems (Gorban et al, 2010). Here I will focus on describing some applications of these approaches to two problems: modeling NFkB pathway and its connection to apoptotic signaling and modeling the mechanisms of miRNA action on protein translation.

## *2.2 Mathematical modeling of NFKB and apoptosis signaling*

The process of induction of the NFkB transcription factor by tumour necrosis factor (TNF) or other ligands together with the interplay between NFkB and apoptotic pathways have been the subject of intensive mathematical modeling during last 20 years. The NFkB pathway is one of the most mathematically modeled with more than 30 mathematical models devoted to various aspects of this signalling (Cheong et al, 2008).

Several attempts were made in order to develop a consensus combined model of NFkB with little success to our knowledge so far. One of the obstacles is incompatibility of various models on the level of their wiring, parameterisation and the level of complexity. In (Radulescu et al, 2008), we developed the most detailed at that time model of NFkB signalling containing 39 chemical species and 65 reactions, described by 90 parameters, and the way to systematically reduce its complexity was suggested. This allowed comparing of the most detailed model to several existing models in a uniform fashion.

In particular, we compared our model to the existing model of NFkB signaling developed by Lipniacki et al, 2004, and showed that the principal difference between two models was in the way the Ikb$\alpha$ gene is regulated in the nucleus. From the complex model both nuclear Ikb$\alpha$ and NFkB complex regulates transcription while the role of the nuclear Ikb$\alpha$ is completely neglected in the Lipniacki's model, which drastically reduced the model's complexity. Other differences between ours and the Lipniacki's model can be attributed to different granularity of description ("complicatedness" rather than "complexity").

Moreover, we suggested a model of minimal possible complexity able to reproduce the experimental data (damped oscillations of NFkB localization). It contained only 5 chemical species (Figure 3) and 12 parameters. These were the key model parameters which correspond to some monomials of the parameters in the detailed model (Radulescu et al, 2008).

In a more recent work, we proposed a new approach for model composition based on reducing several models to the same level of complexity and subsequent combining them together (Kutumova et al, 2013). We suggested the following definition of a model of minimal complexity: this model is the simplest one that can be obtained from the initial model using a set of model reduction techniques and still able to approximate well the experimental data. We proposed a strategy for composing the reduced models together.

This approach was tested on two models describing the cell fate decisions between NFkB-mediated cell survival and apoptosis (programmed cell death). We showed that the reduced models lead to the same dynamical behavior of observable measures and to the *same predictions* as the models-precursors. The composite model was able to recapitulate several experimental datasets, which were used to calibrate the original models separately, but also had new dynamical properties which allowed us to formulate new verifiable predictions.

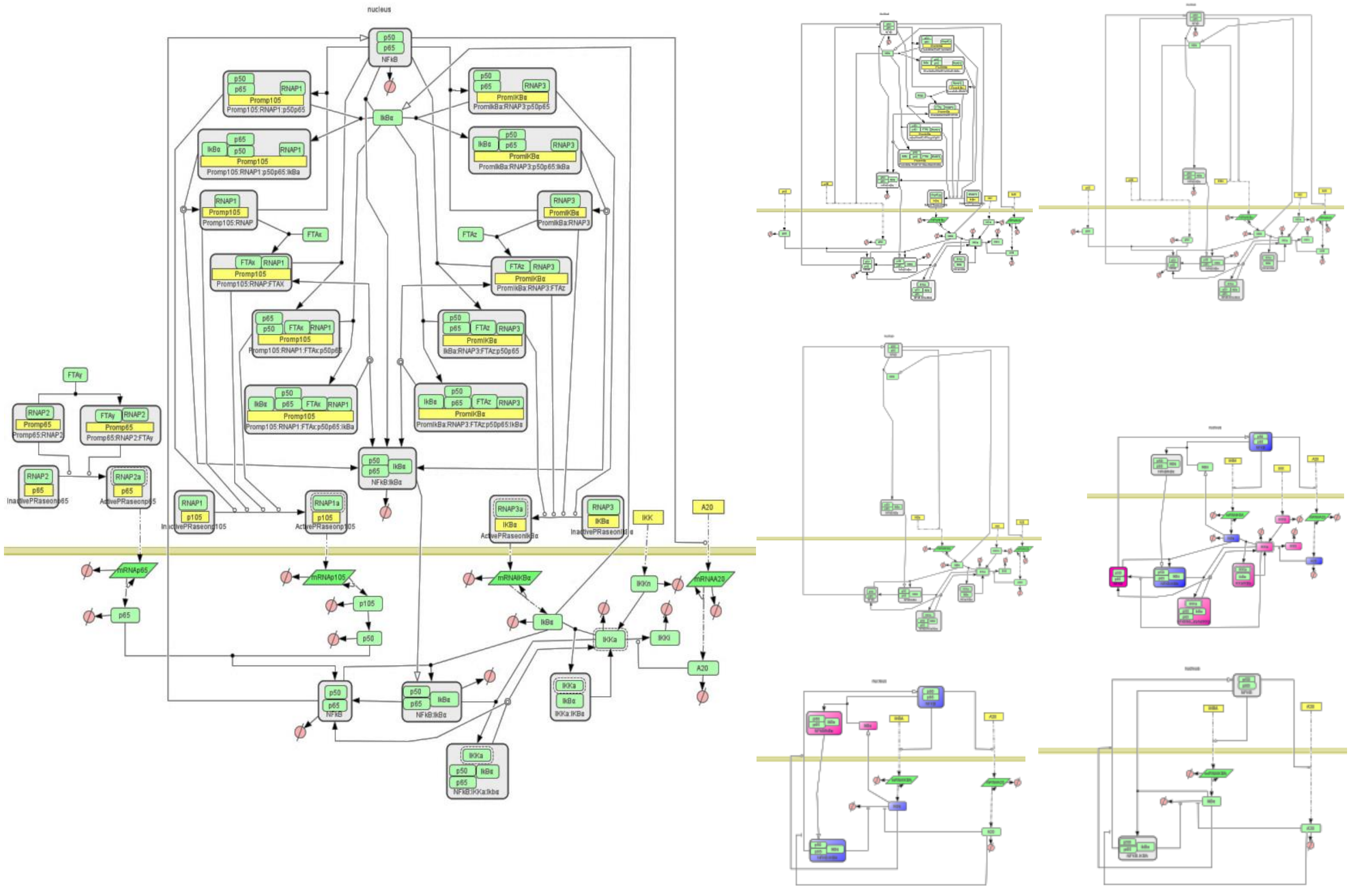

Figure 3. Model reduction of the comprehensive model of NFkB pathway. Left: the initial complex model, containing 39 chemical species and 65 reactions. Right: consequent steps of model reduction. The forth step (second row, left) corresponds to the Lipniacki’s model (Lipniacki et al, 2004), containing 14 chemical species. The minimal complexity model contains 5 chemical species and 8 reactions. Reproduced from (Radulescu et al, 2008).

## *2.3 Mathematical modeling of miRNA-mediated translation repression*

In a series of works (Zinovyev et al, 2010; Morozova et al., 2012; Zinovyev et al., 2013A), I applied asymptotology and model reduction for dissecting complexity of a complex molecular mechanism of miRNA-mediated translation repression.

MicroRNAs can affect the protein translation using *nine mechanistically different mechanisms*, including repression of initiation and degradation of the transcript (Morozova et al, 2012). There is a hot debate in the current literature about which mechanism and in which situations has a dominant role in living cells. The worst, same experimental systems dealing with the same pairs of mRNA and miRNA can provide ambiguous evidences about which is the actual mechanism of translation repression observed in the experiment. It happened that mathematical modeling is able to suggest explanation to existing controversies in the field.

To understand the effect of miRNA on translation, one needs to construct a kinetic model of translation. Two simple models of translation were developed in (Nissan and Parker, 2008), and we started from the detailed analysis of them.

The first model represents a simple cycle of three reactions (Nissan and Parker, 2008; Zinovyev et al, 2010): 1) free ribosomal subunit 40S binds to mRNA, 2) full ribosome is assembled at the start codon, 3) mRNA is translated and ribosomes are released from mRNA. Even this simple model suggests that depending on which step of translation is limiting, the effect of miRNA can be detectable or not detectable. Thus, if 40S binding to mRNA is a rate-limiting step, but miRNA modulates the ribosome assembling step, then the effect of miRNA will not be observed in the experiment (for realistic values of the strength of the modulation). This indeed happens when mRNA with a modified artificial cap structure is used, which makes the initiation step very inefficient (hence, rate-limiting).

In a Science paper (Mathonnet et al, 2007), a wrong conclusion was made that the absence of miRNA effect on the *steady-state rate of protein synthesis*, for mRNA with a modified cap structure, proves that miRNA action requires the normal mRNA cap. Mathematical model shows that it is not possible to distinguish between two explanations: 1) miRNA action is cap-dependent or 2) miRNA acts at the ribosome assembly step (which is not rate-limiting). However, as it was suggested by us, it is possible to distinguish between two mechanisms, if together with the change in the steady-state rate of protein synthesis one would measure also the *relaxation time*, i.e. the time needed to arrive to the new protein synthesis steady rate (Zinovyev et al, 2010).

The second model suggested in (Nissan and Parker, 2008) is non-linear because it explicitly takes into account the turnover of translation initiation factors and ribosomes. It would be natural to determine which reaction in this model is rate-limiting, but this is not possible. Actually, it is known that the notion of rate-limiting reaction step is applicable to only very simple systems, while in the non-linear case the rate limiting “place” can be not a single reaction and, moreover, can change with time. Thus, in complex and non-linear systems the notion of rate-limiting step should be replaced by the notion of *the dominant system* (Gorban et al, 2010). For the non-linear model of translation suggested in (Nissan and Parker, 2008) we performed a complete asymptotological analysis as it is depicted in the previous section. We listed all possible asymptotic system behaviors, and showed how the asymptotic solutions (obtained by the dominant systems integration) change each other in time. As a result, a semi-analytical solution for a non-linear model of translation was obtained and analyzed for possible measurable effects of miRNA (Zinovyev et al, 2010). As a result, we suggested experimental designs in which several mechanisms of miRNA action can be distinguished.

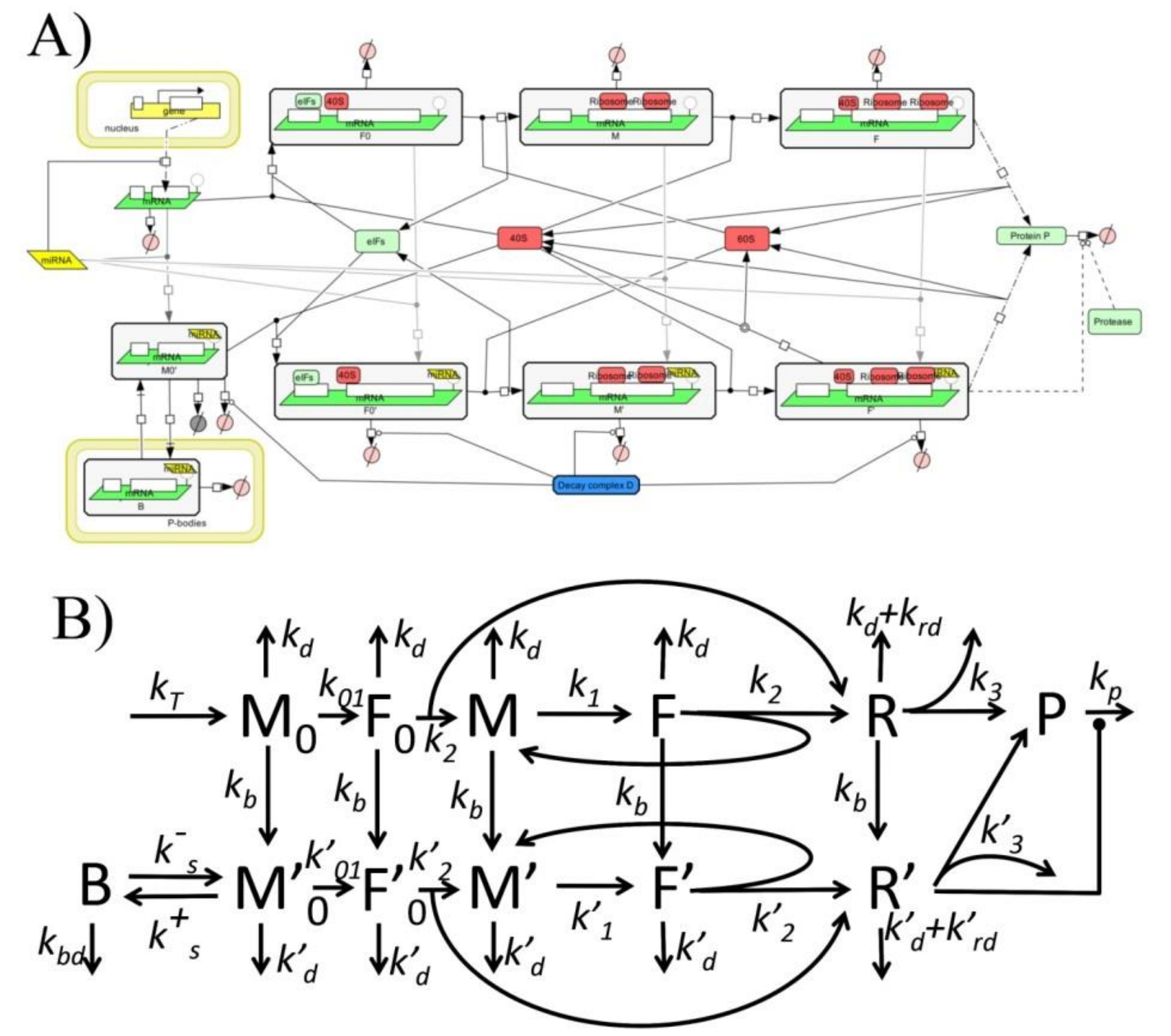

Figure 4. Mathematical model of miRNA-mediated mechanisms of translation repression. This model combines 9 known mechanisms in one reaction network. A) SBGN-based representation of the model. B) Schematic view of the reaction network where M0 is newly produced mRNA, F0 – mRNA initiated with 40S ribosome component for the first round of translation, M and F – are amounts of translated mRNAs without or with 40S ribosomal component sitting at the translation initiation site, R – is the total amount of translating ribosomes and P is the amount of protein. Prime symbol denotes the corresponding states of mRNA with miRNA bound. The derivation of the model is described in (Gorban et al, 2013).

This work was finalized by creating a new kinetic model of coupled transcription, translation and degradation. We build such a model by lumping multiple states of translated mRNA into few dynamical variables and introducing a pool of translating ribosomes (Gorban et al, 2013). In this model, it is possible to simulate all nine known mechanisms of miRNA action (Figure 4). Moreover, it is possible to consider the situation when several mechanisms of miRNA action are present simultaneously, and to predict the measurable effect in this situation.

We dissected the complexity of our translation model by applying the asymptotology approach. We found out that the model has 6 distinct asymptotic behaviors. Some of them correspond to single mechanism of miRNA action, but most correspond to several mechanisms simultaneously. Classification of miRNA action into 6 dynamical types is more constructive from the point of view of observable variables (such as total mRNA and protein amounts, and the polysome profiles) than considering the individual molecular mechanisms. This classification shed light onto the controversies existing in interpreting the results of experiments. Indeed, as our analysis showed, the same mechanism of miRNA action can lead to different translation dynamics depending on the concrete distribution of parameter values and to confusion in the interpretation.

Based on our analysis, we suggested precise recipes (*kinetic signatures*) on how to distinguish between different mechanisms of miRNA action from experiments on observing the dynamics of the amounts of mRNA, protein and the average number of ribosomes translating one mRNA (Morozova et al, 2012). These theoretical results await experimental validation.

In conclusion, studying the mathematical models of complex molecular mechanism of translation showed that the asymptotology approach is an efficient tool for dissecting the complexity into relatively simple scenario, each of which can be analyzed and understood unlike the comprehensive model: in other words, *fighting with complexity of mathematical models is possible*.

# 3. THE MYSTERIOUS COMPLEX GENOME

## 3.1 What we can learn from genome: from genetic code to lifestyle

My first scientific passion in bioinformatics was trying to understand the properties and evolution of the genomes and how they are affected by environment and history. Today we know many genomes with high-precision: genomic sequences are the most reliable experimental data in biology. We can now compare genomes of various species, of individuals of the same species, even of different individual cells in the same biological organism. And still the genomic sequence appears for us as a book written in unknown language, even for very simple organisms. We can distinguish some words, we know much less the genomic “grammar rules”, and we know very little how to read the information about the organism itself and its environment encoded into the genome. Thus, there exists a need in methods for fighting with genome complexity.

In the very beginning of my acquaintance with genomes, I reasoned in a very simplistic science-fiction-like fashion. Let us imagine that there exists a dead planet populated by robots who are equipped with very powerful technological devices and computational methods but do not have any idea about biology. Imagine that they send a space probe to Earth which brings few samples containing some biological material. Very quickly, the robots will detect presence of long molecules of DNA, apparently containing some information. The question is how to read it and find the way to decode it.

One of the natural approaches to do it is the following (Zinovyev et al, 2003; Gorban et al, 2003; Gorban et al 2005B, 2005C; Gorban and Zinovyev, 2007). Let us cut the genome into fragments of approximately same lengths, and characterize each fragment by a vector of frequencies of short words. Thus, a fragment can be characterized by a 4-dimensional vector of single nucleotide frequencies, 16-dimensional vector of dinucleotide frequencies, 64-dimensional vector of triplets, and so on. In addition, *k*-mers can be counted with overlaps or sequentially.

I showed that the unsupervised analysis of these data allows quick discovering that the genome contain relatively short messages (words, i.e. genes) encoded by non-overlapping triplet (letter, i.e. codon) code, separated by less structured spaces. This can be done by studying the 7-cluster structure described below. By comparing many similar genomic sequences it is possible to decode the genetic code and the 20-amino acids alphabet. It is relatively easy to see that the amino acids and the words can be classified into two major classes (which, in biological language, correspond to hydrophilic and hydrophobic proteins and amino acids). From this observation, one could guess the membrane-based organization of life, i.e. existence of cells, in a water-containing environment. It is possible to determine some more specific classes of words: some of them will look like being “pushed” or selected for a particular use of the synonymous letters/codons (corresponding to ribosomes and heat shock proteins), some of them will look like being borrowed from other biological species (corresponding to horizontally transferred genes). It will be possible to estimate the strength of selection, and classify the selection into several types. From the strength and type of selection, one might distinguish fast and slow-growing organisms. Some of the identified factors will happen to be connected to temperature and oxygen presence, thus, it will be possible to assume that some organisms grow faster only at a certain temperature or oxygen concentrations. This investigation program can be continued.

This imaginary exercise is not useless: it does allow to reverse-engineer some organism properties from their genomes. The distribution of genomic fragments, decomposed into non-overlapping triplets, forms the 7-cluster structure in the 64-dimensional triplet frequency space (Figure 5). This cluster structure can be used for unsupervised gene identification (Zinovyev et al, 2003). I studied the properties of this cluster structure in microbial genomes and found out that it can be characterized by four possible symmetry types (Gorban et al, 2005B; Gorban et al, 2005C; Gorban and Zinovyev, 2007). Also, we showed that the simplest context-independent model describe the codon usage with surprisingly high-precision, and that the codon usage is a multi-linear function of GC-content, but this function is different for archaea and bacteria (Gorban et al, 2005B; Gorban and Zinovyev, 2007).

In parallel to studying the basic properties of the 7-cluster structure of the genome, I looked at the properties of codon usage bias. To study the effect of translational bias on gene expression, Sharp and Li, 1987 suggested to associate to each gene of a given genome a numerical value, called Codon Adaptation Index (CAI) which expresses its synonymous codon bias. The idea is to compute a weight for each codon from its frequency within a chosen small pool of highly expressed genes *S*, and combine these weights to define the CAI(g) value of each gene *g* in the genome. For Sharp and Li, the hypothesis driving the choice of *S* is that, for certain organisms, highly expressed genes in the cell have the largest codon bias, and these genes, made out of frequent codons, are representative for the bias. Based on this rationale, one can select a pool of ribosomal proteins, elongation factors, proteins involved in glycolysis, possibly histone proteins (in eukaryotes) and outer membrane proteins (in prokaryotes) or other selections from known highly expressed genes, to form the representative set *S*. Therefore, computation of CAI was based on an expert opinion and also not applicable for those organisms where there is no translational codon bias (for example, slowly growing *Helicobacter pylori*).

We showed that the set *S* can be found in a completely unsupervised manner, by a simple iterative strategy (Figure 5). From the analysis of the properties of the genes in the set, one can tell 1) which type of codon bias is dominating for this genome; 2) tell if the organism is slow or fast-growing; 3) for the fast-growing organisms, predict the average trend of gene expression for the majority of genes. Our approach worked even for relatively complex organisms such as drosophila. The method attracted attention of the scientific community: for example, it found applications for optimization of codon usage in order to increase production of certain proteins in biological experiments or bioreactors.

We designed a computational method for quantifying the codon bias in microbial and eukaryotic organisms. We found out that codon usage can be affected by several types of codon bias (translational, leading-lagging strand bias, GC-content bias, GC3 bias, bias from horizontal transfer). We provided a completely unsupervised computational procedure for computing Codon Adaptation Index (CAI) for the dominating codon bias (Carbone et al, 2003).

Using CAI approach, we studied the genomic properties of all microbial genomes that were available in 2005 (Carbone et al, 2005). We classified them by the type of codon bias and demonstrated that codon bias space reflect the geometry of a prokaryotic ''physiology space.'' In the codon bias space, we identified sets of preferred codons to discriminate: translationally biased (hyper)thermophiles from mesophiles, and organisms with different respiratory characteristics, aerobic, anaerobic, facultative aerobic and facultative anaerobic. This type of approach is particularly interesting for studying those microbes which cannot be cultivated in the laboratory conditions and not directly observed in their natural environment, or for modern *metagenomics* studies.

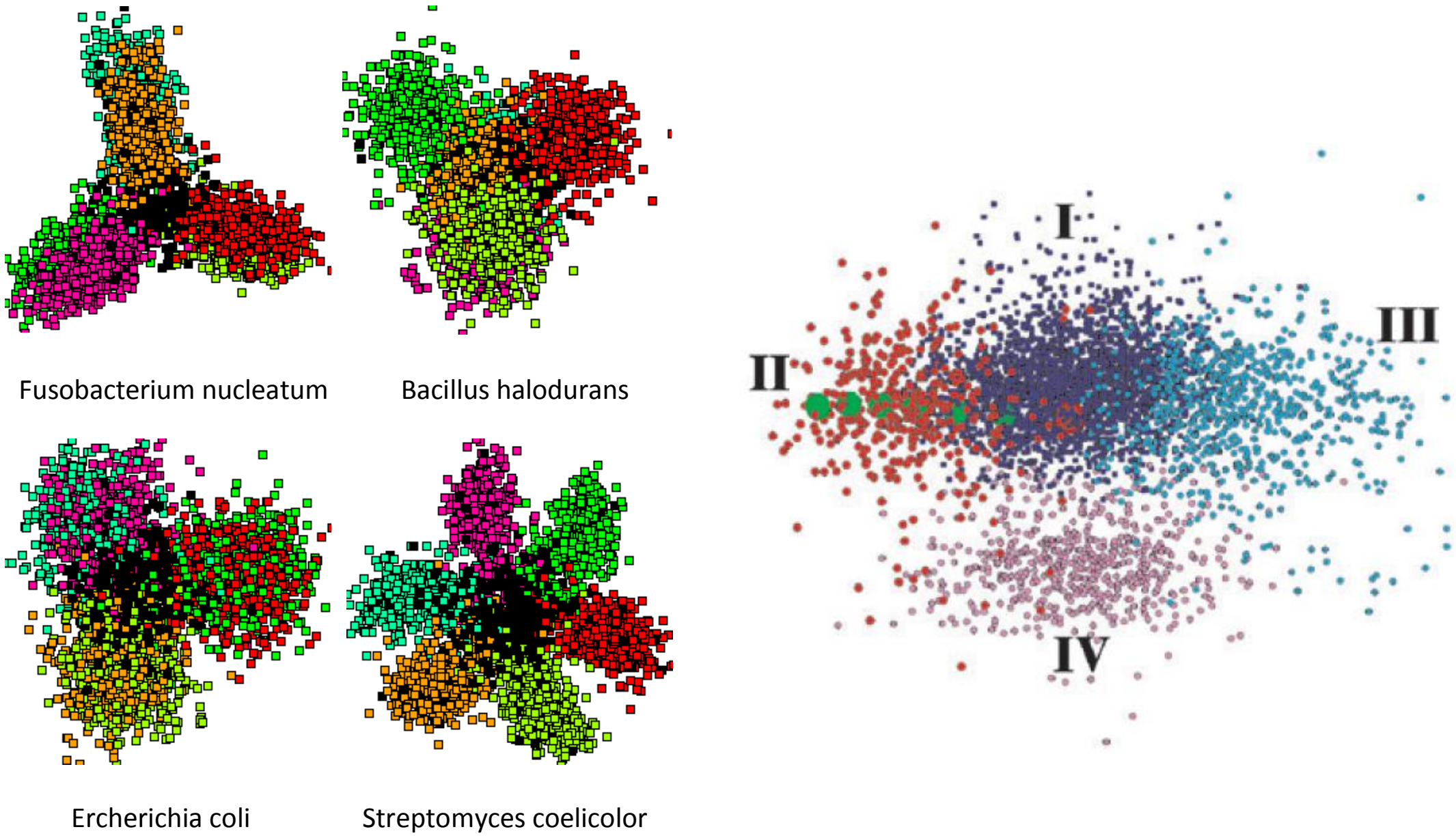

Figure 5. Cluster structures existing in the 64-dimensional space representing genomes as a set of fragments characterized by the frequencies of non-overlapping triplets. Left: the 7-cluster structure corresponding to the presence of coding information in 6 possible frames (color points) and a non-coding fragments (black points). Four possible symmetry types for this structure are shown (reproduced from (Gorban et al, 2005B)). Right: cluster structure corresponding to the distribution of genes (characterized by their codon-usage vectors) in a genome of *E.Coli*. Four clusters correspond to separation into hydrophobic vs hydrophilic (cluster IV vs I+II+III), a set of highly translated genes (I) and a set of horizontally transferred genes (III). Green circles shows several iterations of the algorithm to automatically find the set of the "most biased" genes (reproduced from (Carbone et al, 2003)).

## *3.2 Genome functional fraction estimates*

Recent "RNA revolution" in molecular biology showed that the coding part of genome explains only a fraction of organismal complexity (Taft et al, 2007). The vast genome fraction which is for long was considered as "junk" or "dark matter" of DNA contains a plethora of information absolutely required for cell functioning. However, the question "how big is the functional genome part" remains open and hotly debated (Ponting and Hardison, 2011). It is also a part of an old discussion on how is the length of non-protein-coding DNA related to eukaryotic complexity.

Now it is clear that most of genome is transcribed, but it might be that this transcription is "pervasive" and serves for creating rather a "bouillon" of possible molecular structures, not necessarily having clear biological function (Clark et al, 2011). Experimental investigations of the functional fraction of the genome inevitably face the conceptual difficulty of defining what a biological function is. In the theoretical studies, the functional fraction is frequently associated with that part of the genome which is affected by positive purified selection: hence, this part tends to be conserved in evolution unlike the non-functional part.

Historically, the first estimates of the genome functional fraction size came from the genome-wide alignment studies. Thus, comparing human and mouse genomes gave an estimate of 5% (including the coding part) for the functional portion of the human genome (that one which is under purifying selection) (Chinwalla et al, 2002). However, a careful analysis of this alignment showed substantial methodological controversies. The key parameter of the genome-wide alignment is the length *W* of the widow size used for local alignment. In that study, this parameter was arbitrarily set to *W* = 50. Changing the value of this parameter can change the final functional fraction estimate to any number from 3% to 8% (Chiaromonte et al, 2003). Moreover, the dependence of the estimate is not a continuous function of *W*, so no distinguished value of *W* exists.

Since then, this estimate was computed many times: there exist about 17 papers devoted to this question (Ponting and Hardison, 2011), which give estimates varying from 3% to 15%. Importantly, most of them were done based on genomic sequence analysis, and the notion of the functional fraction was a synonym of the *fraction of the genome under purifying selection*.

In 2007, together with my colleagues we made an attempt to estimate the functional fraction of the human genome from completely different theoretical perspective, independent on genome alignment-based estimates. As a result, we published a paper which was called “How much non-coding DNA do eukaryotes require?” (Ahnert et al, 2008). Our estimate is based on assuming a particular type of the scaling law for the size of the regulatory network necessary to support the growing number of protein-coding genes.

In prokaryotes, a quadratic relationship between the number of regulatory genes and the total gene number has been demonstrated in the literature (Croft et al., 2003). This leads to a “complexity ceiling” in the maximal number of genes in a prokaryotic genome (the maximal size of prokaryotic genome is about 12Mbases). We followed an assumption that eukaryotes bypassed this complexity barrier by exploiting cheaper RNA-based regulatory mechanisms which are encoded in non-protein-coding part of DNA. We hypothesized that the growth of the total regulatory network remains quadratic with respect to the total number of genes. Using available data on eukaryotic genomes and the sizes of their protein-coding parts, we evaluated the human genome functional fraction as 7%. Future functional studies will show if this estimate is realistic: however, it is convergent with an average 6.5% value over all published sequence alignment-based estimates (Ponting and Hardison, 2011). Together with this, we made estimates for genome functional fraction of many other organisms (for example, 34% for *Drosophila melanogaster* and 55% for *Caenorhabiditis elegans*).

## 3.3 Mysteries of cancer genome

My current interest to genome properties focuses on the problem of *cancer genome evolution*. Currently we know well that most of progressing tumors are characterized by one or several periods of genomic instability (Loeb, 2011), which are manifested in the genomic profiles of tumor cells by presence of many types mutations of different scales. These mutations include point-wise and small scale mutations, local changes in the gene copy numbers, as well as global genome rearrangements, including aneuploidy and genomic translocations. Whole-genome duplication events are not unusual in the evolution of cancer cells which is followed by massive loss of redundant gene copies.

There are many questions that can be raised looking at the “snapshot” of genome evolution that we observe in a tumoral biopsy. Among them, reconstruction of the history of large-scale genomic events and characterizing quantitatively the mutation spectra are intensively studied. My interest is focused on the question of *how these genomic modifications are constrained by the properties of the protein-protein interaction networks*. This is a relatively poorly studied question in the current literature, though it is well-known that changes in gene and chromosome copy numbers significantly affect the functioning of biological networks. Changes in gene copy numbers affect gene expression, and, therefore, violate protein dosage balance, which in turn leads to malformation of essential protein complexes or dysregulation of essential cellular pathways. Requirement of 1:1 ratio of the components of Mad1:Mad2 complex in yeast is one of the examples of such dosage balance control which is essential for stringent mitotic checkpoint surveillance mechanism that monitors unattached or improperly attached kinetochoresis (Potapova et al, 2013).

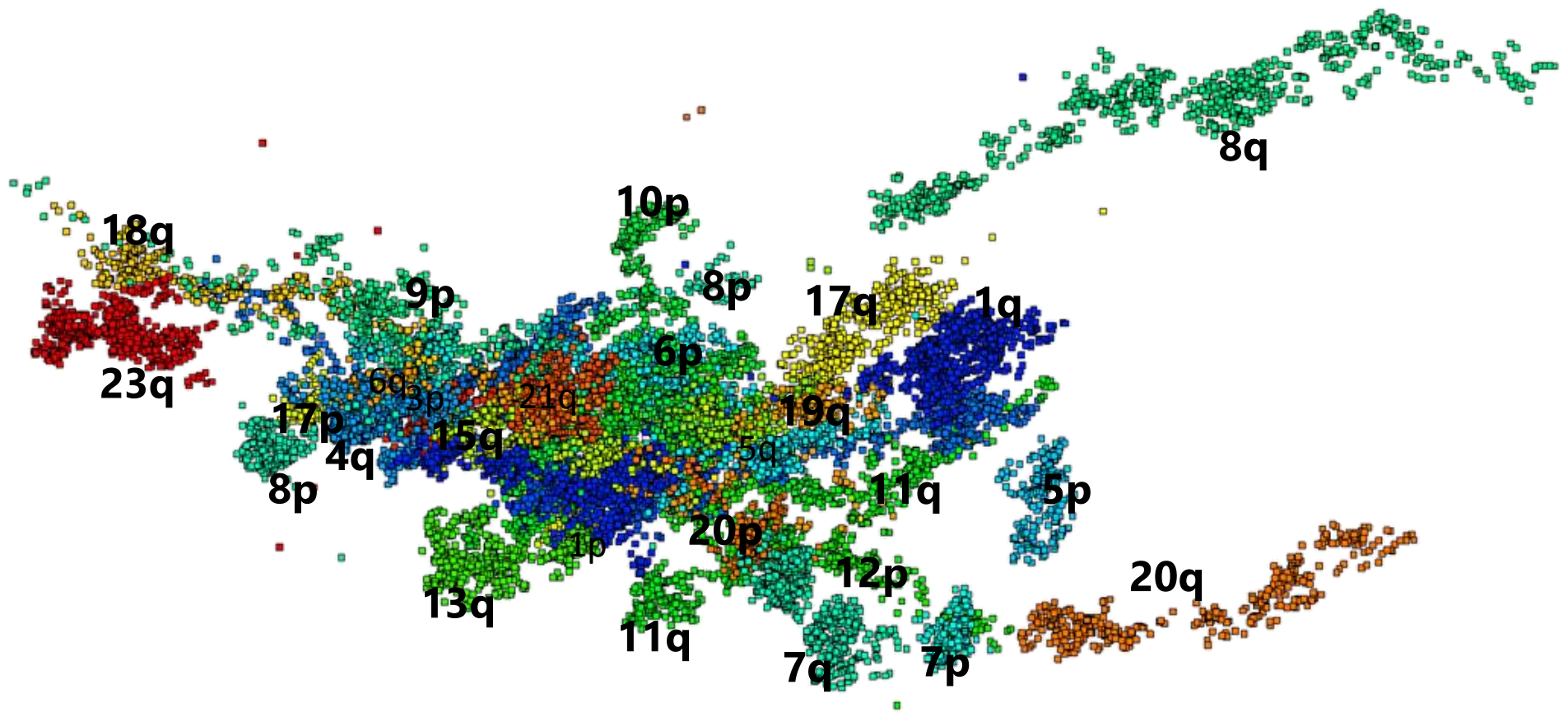

Figure 6. Visualization of cancer genome. This is a PCA plot of genome locuses characterized by their profiles of copy number changes in a series of 160 breast and ovarian cancer cell lines, available in Broad Cancer Cell Line Encyclopedia (http://www.broadinstitute.org/ccle/home). Different colors designate different chromosomes (starting from blue, 1st chromosome, to red, 23rd, or X-chromosome). The horizontal PCA axis is approximately associated with the frequency of gains (on the right) and losses (on the left). The vertical PCA axis does not have evident interpretation. Those chromosome arms located closely on the plot have similar profiles of gains/and losses. With some imagination one can recognize in this picture an image of a shellfish with two big claws. The scientific challenge here is to associate this structure with some features of the genome-wide reconstructions of the protein-protein interaction network.

Even before starting to study this question, one should resolve technical problems connected with extracting the reliable estimates on gene gains and losses from the raw data. This is a simple problem for those cancer genomes which are relatively little rearranged, and rather complex for those genome having massive modifications. The situation is complicated by the fact that the cancer genome could undergo tetraploidization and that the control material representing normal cells is frequently absent. Moreover, tumor sample usually contains a significant proportion of normal cells and characterized by genomic heterogeneity (co-existence of several subclones). I participated in creating software having the purpose of quantifying gene gains and losses in cancer genomes (Boeva et al, 2011).

Currently I work on quantifying constraints on the evolution of cancer genome coming from the structure of global biological networks (in particular, their compositions of molecular complexes). For this purpose we use publicly available data on cancer genomic profiles (Figure 6), and highly-curated pathway database, constructed by our team (Atlas of Cancer Signaling Networks, http://acsn.curie.fr ).

# 4. CANCER AS A COMPLEX SYSTEM

I study cancer by computational approaches for almost 9 years now. My experience together with experience of some of my colleagues was recently summarized in the book "Computational Systems Biology of Cancer" which is one of the first textbooks in the field (Barillot et al, 2012). One important thing to say about studying cancer is: *to understand cancer genesis and progression, one has to understand functioning of many (if not most) normal cellular functions*. This is a common spirit of the research in my institution, Institut Curie. The more we understand cancer, the more we realize how systemic it is as a disease. Dysregulation of molecular mechanisms leading to cancer concerns various processes such as cell cycle, cell death, DNA repair and DNA replication, cell motility and adhesion, cell survival mechanisms, mechanisms of immune system and angiogenesis; and usually most of them are involved in the same tumor. Cancer does not invent completely new molecular mechanisms: it rather hijacks the existing molecular programs (of inflammation, development, wound healing, and others) in order to insure the tumor growth and dissemination of cancer cells through the body. Hence, we study the molecular biology of cancer as a specific distortion of normal cellular functioning.

For computational systems biology of cancer, this approach dictates the following challenges: 1) represent formally and in sufficient amount of details the existing knowledge about those molecular processes whose involvement in cancer clearly demonstrated (these are recapitulated in the famous cancer hallmarks, Hanahan and Weinberg, 2011); 2) collect and integrate existing quantitative data on cancer genesis and progression and develop methods to analyze them in the light of the knowledge of a normal cell; 3) create mathematical models able to describe distortions of normal cell functioning as a cause of cancer, and to predict the effect of various perturbations; 4) use data analysis and mathematical modeling to suggest new therapeutic options.

Our "Computational Systems Biology of Cancer" (http://sysbio.curie.fr) research group, which I coordinate at Institut Curie, contributes to some extent to all these directions. I will describe briefly some of our achievements below, roughly following the plan outlined in the previous paragraph.

## *4.1 Formalization of knowledge in cancer biology*

Starting from 2006, our group invests a lot of efforts into formal representation of molecular mechanisms involved in cancer. Since then I am deeply involved in this activity.

My source of inspiration was a special graphical language which was suggested by the systems biology community at that time in order to unify and formalize the graphical representation of biological diagrams. This language is called Systems Biology Graphical Notation (SBGN) (Le Novere et al, 2008). Just like the system of graphical symbols used by electrical engineers to depict the design of electrical circuits, SBGN provides symbols and rules for specifying molecular biology entities (such as proteins, genes, RNA molecules and their modifications, etc.) and processes (biochemical reactions, transport reaction, processes of transcription and translation, etc.). Creation of such a language is a great step towards converting the knowledge about biochemical mechanisms, which now exists in millions of human-readable publications, into a computational resource which can be used in analysis of data and biological networks and creation of mathematical models.

My first initiative was designing a comprehensive map of a pathway which is dysregulated in the absolute majority of cancers: RB/E2F pathway. Retinoblastoma protein (RB1) serves as a cell cycle break, providing the checkpoint mechanism of transition between G1 and S phases of the cell cycle. If this break is intact in a cell then the progression through the cell cycle is tightly controlled. The text book description of this pathway is quite simple and tells that RB1 protein normally binds E2F-family transcription factors (E2F1, E2F2, E2F3) unless it is hyperphosphorylated by Cyclin D (CCND1) kinase. The comprehensive description of this pathway is much more complicated (one has to have in mind that there are three distinct pocket proteins in RB family, and RB1 has 16 phosphorylation sites). We have created a comprehensive map of this mechanism describing functioning of 78 proteins present in 208 distinct molecular forms connected by 165 chemical reactions (Calzone et al, 2008). This map recapitulates information from 350 scientific publications in a formal and computer-readable fashion, using CellDesigner software (http://celldesigner.org).

In parallel to constructing the pathway I started developing software for the structural analysis of the pathways created by CellDesigner, called Biological Network Manager (BiNoM) (http://binom.curie.fr) and for visualization of large comprehensive maps, called NaviCell

(http://navicell.curie.fr). With time, in this software we implemented many tools for manipulating biological networks, converting them from one standard format into another, applying some complicated graph theory algorithms (such as solving *the minimal cut set problem*, (Vera-Licona et al, 2013)). This is why BiNoM became a visible tool in the systems biology community (Zinovyev et al, 2008; Bonnet et al., 2013A; Bonnet et al., 2013B).

During several years I participated in creating comprehensive maps of molecular interactions for many cellular processes, related to cancer (maps of cell cycle, DNA repair, apoptosis and mitochondrial metabolism, cell survival, epithelial-to-mesenchymal transition (EMT) and cell motility). These maps already describe functioning of the majority (60%) of cancer driver genes as they are defined recently by (Vogelstein et al, 2013), and we constantly enlarge the maps with new data. Recently, all these pathway maps were collected into a pathway database, Atlas of Cancer Signaling Networks (ACSN), focused on cancer-related mechanisms, which was made available online (https://acsn.curie.fr). The concept of this pathway database is different from the majority of the existing ones because it represents an interactive map of molecular interactions. The whole content of the database is visually presented to the user in the form of a global 'geographic-like' map browsable using the Google Maps engine and semantic zooming. The associated blog provides a forum for commenting and curating the ACSN content. The global ACSN map is hierarchically organised in a set of interconnected maps. Each map is accessible individually, as in a geographical atlas. The database is manually assembled by qualified researchers with participation of the experts in the corresponding fields. The construction of the map involves careful mining of molecular biology literature and additional sources of information, and detailed depicting of molecular mechanisms using a standard representation of biochemical reaction networks.

Formalization of knowledge on the functioning of biological networks is a very important step in fighting with their complexity. Using comprehensive (not over-simplified) reconstructions of cancer-related processes, we can have a realistic idea on the complexity of their structure, and start developing adequate algorithms for quantifying and abstracting this complexity for further use in data analysis and mathematical modeling.

## *4.2 Vertical and horizontal integration of cancer data*

Cancer biology now possesses enormous amount of genome-wide quantitative molecular data. These data can be classified by the type of molecules which are characterized (DNA, RNA, proteins, metabolites, etc.), by the type of measured features (DNA copy number, mutations in exons, DNA methylation, metabolite concentration, protein modifications), by the technology used to obtain the data (genetic chips or sequencing), by the type of the biological samples (cell lines, xenografts, tumor sample, etc.), by the type of the study (tumor profiling, GWAS, drug or siRNA-based screenings, time-resolved experiments on cell lines, etc.). In addition to molecular data, one can have access to large collections of histological section images, collections of fluorescent images, large-scale videomicroscopy, etc. For more details, one can read the corresponding chapters in (Barillot et al, 2012).

However, there are several common characteristics in all these variety. First of all, it is that any particular type of these data alone characterizes the cell only partially and that all these data are not 100% reliable and typically contains many measurement errors. In addition, the molecular data are affected by various sources of bias, connected to experimental protocols of sample preparation and processing of the biological material. One can say that we study the enormous cancerous or normal cells complexity by slicing it; where each slice represents a particular type of molecular information (genome, transcriptome, proteome, metabolome, other types of "omics" data). Therefore, significant efforts are needed to integrate the data together. One can distinguish vertical and horizontal types of data integration (Figure 7). Integration of data is not merely a technical task of putting data together in the same folder or a database; it is much more a scientific problem, because *data integration should bring new emergent knowledge,* which is not possible to obtain from any single dataset alone.

*Vertical data integration* assumes that the same biological object is characterized by multiple data of different types. For example, the same tumor sample can be profiled for gene expression, genome structure, genome methylation, protein concentrations, etc.) The task of vertical data integration consists in assembling a puzzle from all such pieces in such a way that a new knowledge would emerge from such integration. For example, integrating mRNA and miRNA expression allows finding functionally related groups of miRNA which is not possible from miRNome alone.

By contrast, *horizontal data integration* is applicable when the same type of data is collected many times for different sets of objects of the same nature. For example, similar cohorts of cancer patients are profiled for their transcriptomes in different cancer centers worldwide. The task of horizontal data integration consists in improving the reproducibility and significance of the statistical signals. This is not reduced to mechanistic merging the data together: usually, merging data is not a good idea because of the so called *batch effect*. Horizontal data integration, sometimes also called *data meta-analysis*, allows distinguishing between biologically relevant signals and technologically induced or laboratory-specific data artefacts. An example of such meta-analysis was already shown in this manuscript in Figure 2.

A specific and important case of data integration is combining our knowledge of biological networks (represented in some kind of mathematical models, structural or dynamical) and the above described types of high-throughput data. This is a task of outmost importance for systems biology, and it still represents one of the greatest and not completely resolved challenges.

---

During past years, I contributed both for developing methodology of data integration in cancer biology and doing data integration for some specific cancers. In particular, by this approach we studied Ewing sarcoma. I will describe in some details this research project below, but before I would like to summarize my methodological contributions to this field.

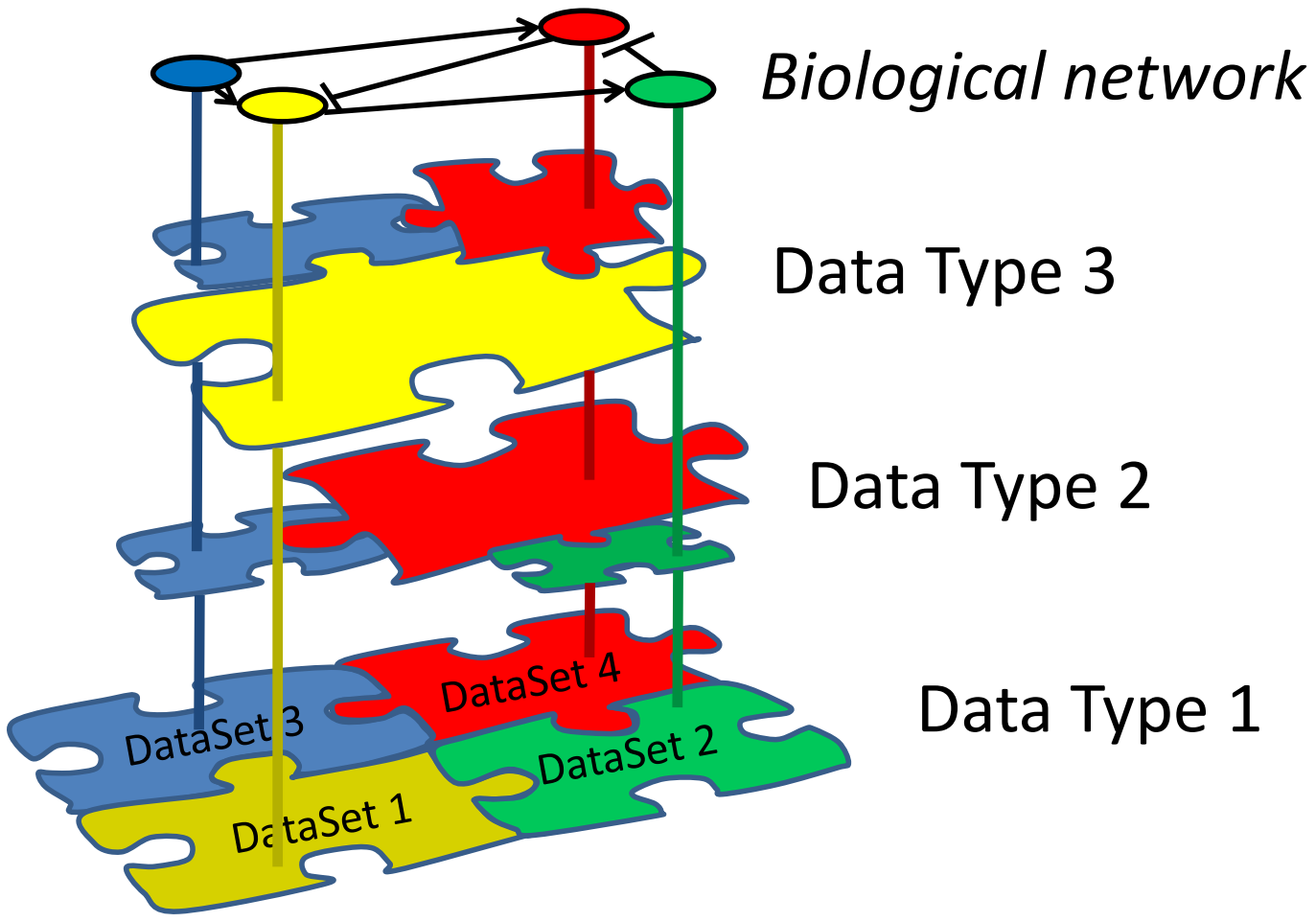

Figure 7. Metaphor of data integration in biology. Vertically, different types of data are combined for the same system in order to obtain new emergent knowledge about it. Horizontally, the same type of data is integrated for different datasets for meta-analysis and improving statistical significance and biological relevance. At the top the data are matched to the biological networks or mathematical models. The color of a puzzle piece signifies different data sets; the size denotes the available volume of data.

We developed a method which can integrate the knowledge on biological network in any type of data analysis, including supervised and unsupervised classification of tumor samples, using their transcriptomes and information on a biological regulatory network (for example, the global network of protein-protein interactions). We suggested performing the data analysis focusing on that component of gene expression profiles which behaves *smoothly* on the graph representing the biological network (Rapaport et al, 2007). This reasoning can be justified by several biologically relevant assumptions (such as assuming the balance of components' stoichiometry in protein complexes). In our study, this approach helped to identify biological processes involved in response of yeast colonies to low-dose irradiation.

Another question on which we have answered using biological networks is as simple as "how many genes one should select from the list of genes, ranked by some measure of significance?" Most of existing methods suggest using a threshold on the statistical significance. However, the value of this threshold is difficult to justify. The commonly used 5% of statistical significance is only a historically accepted number; moreover, it should be adjusted for multiple testing. We suggested to take as many top-ranked differentially expressed genes as needed to obtain an Optimally Functionally Enriched Network (OFTEN) formed by protein-protein interactions (Pinna et al, 2012; Kairov et al, 2012).

---

One of the examples of successful and useful data integration is our long-term project on studying Ewing sarcoma, together with INSERM U830 unit (this project was funded by ANR and INCA French funding institutions in 2006-2012, see http://bioinfo-out.curie.fr/projects/sitcon, and became a part of European Framework 7 large-scale systems biology of cancer program from 2010, http://www.ucd.ie/sbi/asset/ ).

Ewing sarcoma is the second most frequent pediatric bone tumor. In most of the patients, a chromosomal translocation leads to the expression of the EWS-FLI1 chimeric transcription factor that is the major oncogene in this pathology. Relative genetic simplicity of Ewing sarcoma makes its particularly attractive for studying cancer in a systemic manner. Silencing EWS-FLI1 induces cell cycle alteration and ultimately leads to apoptosis, but the exact molecular mechanisms are still unclear.

A large corpus of molecular data on functioning was collected by several laboratories in Europe, including Institut Curie (where the majority of Ewing sarcoma patients are treated) and Children's Cancer Research Institute in Vienna. These data include a) molecular profiling (at the level of genome, transcriptome, miRNome, proteome) of tumor samples, which are annotated with clinical information; b) data on experiments with Ewing sarcoma cell lines, including those inducible systems where the expression of EWS-FLI1 can be controlled; c) siRNA-based screenings with further high-throughput cellular phenotyping, measuring the effect of gene silencing on apoptosis and proliferation; and many others.

In attempt to make sense out of these data, we have developed much know-how on particular aspects of horizontal and vertical data integration. For example, for integrating data on expression of miRNAs and mRNAs in patient samples, we suggested a new method for measuring statistical connection between expression levels, so called "antagonism pattern" (Martignetti et al, 2012), different from the standard correlation approach. Using this and other measures we constructed and analyzed global miRNA-mRNA correlation networks which led to some predictions which we started to validate experimentally. We investigated the DNA-binding properties of EWS-FLI1 by integrating ChIP-Seq and gene expression data in cell lines (Boeva et al, 2009). We applied blind source separation methods for deconvoluting miRNA expression data and for comparing transcriptomes in various pediatric cancers.

But most of the efforts were put for the reconstruction of regulatory network connecting EWS-FLI1 and apoptosis and cell-cycle phenotypes (Stoll et al, 2013), using integration of different sources of data. This network (see Figure 8) was constructed from analyzing transcriptomic time-series measured in the inducible cell lines, using model-based time-series analysis and extracting the information on molecular interactions from biological literature. A part of this network was validated by systematic siRNA-based gene silencing and measuring the changes of gene expression by RT-QPCR in several Ewing tumor cell lines, reverse-engineering of a part of the network from these experimental data, and comparing to the literature-based network. The comparison confirmed the validity of the majority of regulatory links. In addition, the structure of Ewing regulatory network was validated using gene expression data in Ewing sarcoma tumor samples. In particular, we identified several novel direct targets of EWS-FLI1, such as CUL1 gene, which was experimentally confirmed by ChIP and silencing experiments.

Data integration is an important tool in fighting with complexity of biological systems. The methods of data integration allow assembling different slices of a biological system description in one global view, and to distinguish important and reproducible biological signals from those determined by particular technical biases specific for a technology used or to a place where the data are generated and processed.

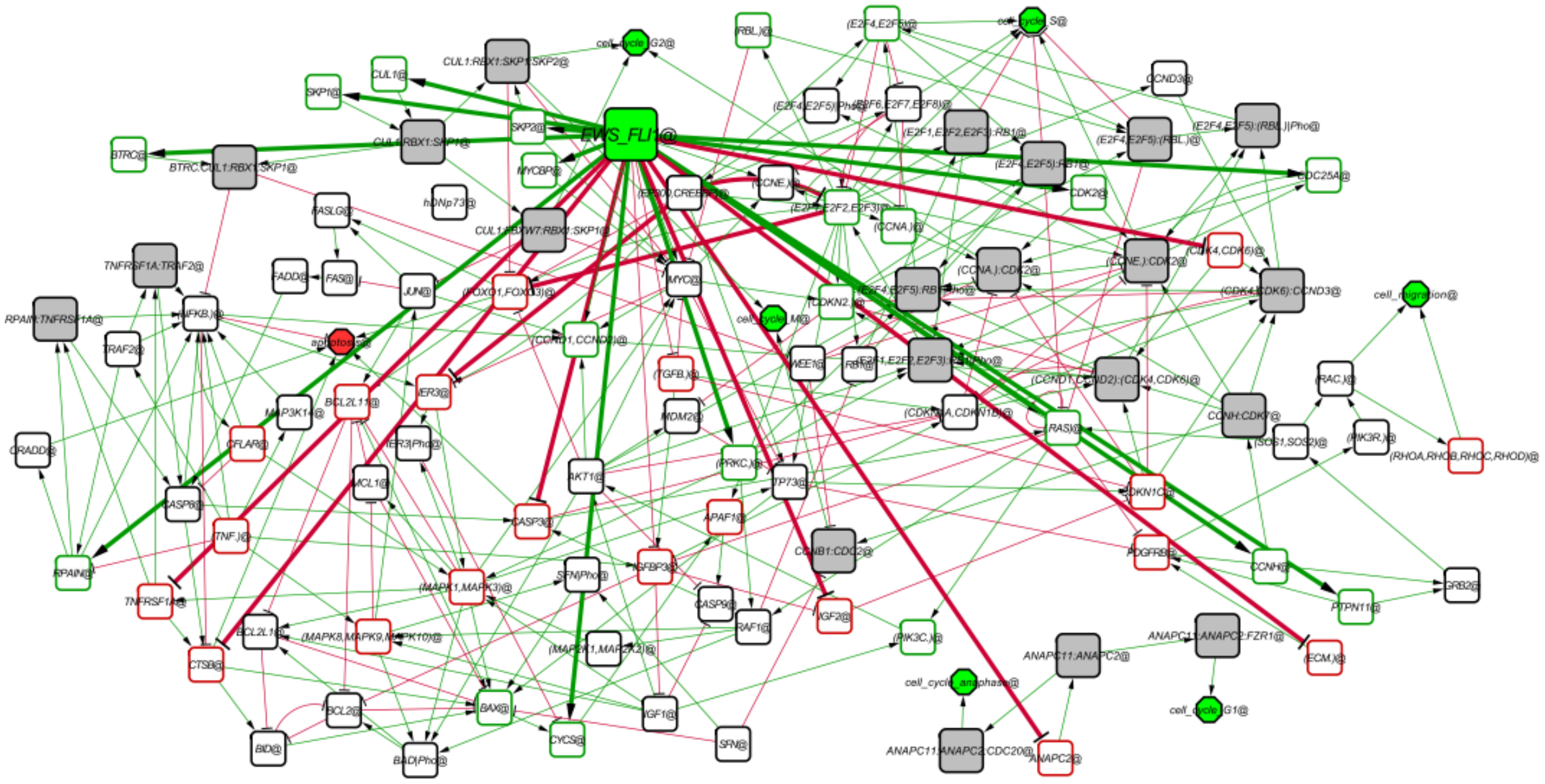

Figure 8. Network of EWS-FLI1 effects on proliferation and apoptosis, derived from analysis of high-throughput data, reading the literature and experiments on gene silencing. White nodes represent genes or proteins, grey nodes represent protein complexes. EWS-FLI1 (green square) and cell cycle phases/apoptosis (octagons) represent the starting point and the outcome phenotypes of the network. Green and red arrows symbolize respectively positive and negative influence. Nodes with green frame are induced by EWS-FLI1 according to time series expression profile and nodes with red frame are repressed. The network structure shows intensive crosstalk between the pathways used for its construction, up to the point that the individual pathways cannot be easily distinguished. The network includes new links inferred from experimental data (thick arrows) from transcriptome time series and siRNA/RT-QPCR. Reproduced from (Stoll et al, 2013).

## *4.3 Cell fate decisions in cancer cells*

One popular Einstein's quote that "Nature integrates empirically" can be repeated with respect to the biological cells too. A biological cell constantly performs complex integration of the signals coming from the environment. This integration is also conditioned on the internal cell state. The result of such "integration" can be "a value" from a discrete set, i.e. a decision to initiate and launch one of the existing cellular programs. For example, a cell can trigger the apoptotic or necroptotic program, i.e. a *program of cellular suicide*, as a response to some unfavorable conditions (e.g., irreparable DNA damage or starvation) or as a part of the normal development program. Another example of such a decision – triggering Epithelial-To-Mesenchymal Transition (EMT) - is a requirement for cellular motility. Let us call *cell fates* triggering such programs (some of them are irreversible), and *cell fate decision* process the process of "empirical integration".

The cell fate decisions are computed by the cells through complex signal transduction pathways and their interactions. The design of these pathways is a result of optimization by natural selection of two mutually opposite requirements: *a possibility to control the decisions*, and *robust functioning, i.e. resistance to various perturbations*, including mutations in the pathway genes (Barillot et al, 2012). Certain mutations or combinations of them can drastically change the cell fate decision process which can lead to various diseases.

One of the most crucial cell fate decision processes is a decision between cell survival and death. Tumour initiation and progression is characterised by a violation of a balance between cell survival and cell death, which is tightly controlled in normal conditions, towards excessive survival in cancer. The balance is normally ensured by a system of molecular switches that trigger irreversible cellular decisions for a particular cell fate in some particular conditions. Compromising the normal function of these switches leads to carcinogenesis and other systemic diseases.

For systems biology of cancer, a very important task is to reveal the logics of the natural empirical integrations, and reproduce it, to some extent, on a computer in the form of mathematical models. We found out that it is convenient to describe the process of cell fate decisions by discrete (logical) modeling which has been developed in computational biology since 1960s. The process of cellular fate decisions in this approach is recapitulated in the form of a state transition graph, describing a finite number of cellular states and transitions between them characterized by certain probabilities. Some of the cellular states represent fixed points, and they are associated with cell fates (or phenotypes). Studying random walks along the state transition graphs, it is possible to estimate the probability of reaching particular fixed points which can be interpreted as a probability to observe a particular cell fate in the biological experiment. The state transition graphs are usually represented in a compact fashion using a biological regulatory network (a directed graph). The nodes (molecules, their modifications, or molecular processes) in this network are assumed to be characterized by a discrete number (state), accompanied by the rules of updating this number based on the states of their immediate upstream neighbors (regulators). In the simplest case, the state is "0" or "1", representing an "active" or "inactive" state.

We used this approach to recapitulate the process of survival-cell death decisions in a well-known experiment where the cells are exposed to a concentration of TNF of FASL proteins (Calzone et al, 2010). As a result, the cells can induce necrotic or apoptotic programmed cell death or survive through induction of NFkB pathway (see Figure 9). The most reliable knowledge about functioning of the cell fate decision molecular machinery was assembled, using GINsim software (http://ginsim.org/), in the form of a regulatory network (Figure 9a), where the nodes represents proteins or their modifications, small molecules (such as ROS) and molecular processes (such as Mitochondrial Permeability Transition (MPT)). Each node was accompanied by an update rule (for example, how BAX protein changes its state depending on the states of CASP8 and BCL2). The state transition graph was computed and analyzed for probability of observing Survival, Necrosis, Apoptosis cell fates in the wild-type cell, and in all single mutants where the state of certain proteins was fixed to "inactive" or "active" states (Figure 9b).

These predictions were systematically compared with the experimental data of the cell death phenotype modifications observed in various mutant experimental systems, including cell cultures and mice (Figure 9c, Calzone et al, 2010). The model was able to qualitatively recapitulate all of them and to suggest some new yet unexplored experimentally mutant phenotypes. The most interesting in this setting would be to consider synthetic interactions between individual mutants, when several nodes on the diagram are affected by a mutation.

Together with this, we developed a toolbox of methods to analyze the properties of the state transition graph. For example, we were able to predict cell fate decisions after transient exposure to ligands. We tested importance of the "switching speed" of certain proteins and importance of individual regulatory links for the process of cell fate decision (Zinovyev et al, 2010). We used specific model reduction techniques to reduce the model to a minimal configuration which revealed the pattern of its organization: "a tristable switch" (Figure 9d, Calzone et al, 2010; Zinovyev et al, 2012; Calzone et al, 2012).

In another project, we used similar approach to predict the effect of mutations in the regulatory mechanism, controlling the switch between Epithelial and Mesenchymal cell phenotypes. This switch (EMT) is associated with appearance of metastases in cancer, because it allows the cells to detach from the primary tumor and become mobile in order to arrive to the distant organs (Valastyan and Weinberg, 2011). The regulatory mechanism was assembled from the P53, WNT and NOTCH pathways whose involvement in EMT regulation was demonstrated (the description is available at https://navicell.curie.fr/navicell/maps/notchp53wnt/master/ ). We predicted that a combination of P53 knock-out and overexpression of NOTCH would induce EMT and not the single mutations alone. This prediction was validated in a genetically modified P53-null mouse model of intestinal cancer, where, indeed, conditional overexpression of NOTCH lead induction of EMT at the front of the primary tumor and to rapid and early metastasation, which was not possible to achieve before (Chanrion et al, 2013).

## *4.4 Exploiting synthetic lethality: a new hope in cancer treatment*

In the beginning of the 20th century, Dr. Paul Erlich suggested that for treating cancer one has to develop a drug which would serve as a "magic bullet". The "magics" is manifested in the fact that the ideal cancer drug should selectively kill cancer cells without affecting normal cells. One hundred years took to start exploiting this concept in clinical practice. For example, imatinib (Gleevec) drug specifically targets only those blood cells containing the translocation producing the chimeric BCR-ABL gene in chronic myelogenous leukemia.

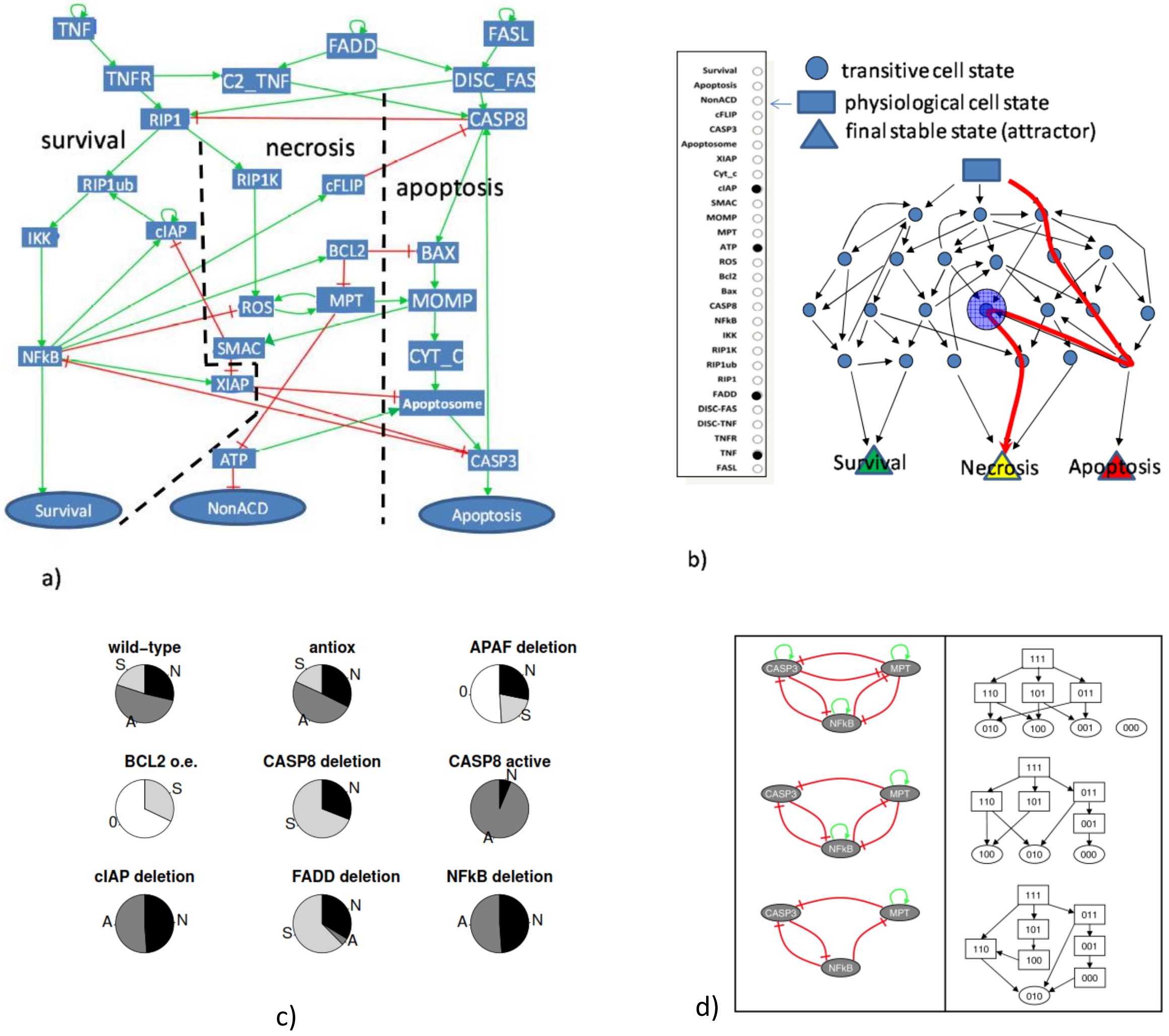

Figure 9. Logical model of cell fate decisions between survival, apoptosis and programmed necrosis (necroptosis). Here ”A“ denotes Apoptosis, ”N“ denotes Necrosis and ”S“ denotes Survival, ”0” denotes Naive state, in which the cell is not inivolved in the cell fate decision process. a) The diagram describing the cell fate decision mechanisms; b) cartoon illustration of a state transition graph on how the probabilities of different cell fates were computed; c) distribution of probabilities of cell fates for some mutant simulations; d) “tristable” trigger patter of the cell fate decision mechanism (top row) with the corresponding state transition graph (on the right) together with various mutation types. Taken from (Calzone et al, 2010; Zinovyev et al, 2012).

Since 1990s, a novel approach for developing the magic bullet was suggested which consists in exploiting known synthetic lethal genetic interactions. Synthetic lethality is a phenomenon when two non-essential for cell viability genes *A* and *B* give lethal phenotype when knocked-out simultaneously. The simplest mechanistical paradigm explaining synthetic lethality is that genes *A* and *B* participates in two redundant pathways performing the same essential function.

If a cancer cell contains a loss-of-function mutation in gene *A* then knocking-down the activity of its synthetic lethal partner *B* will specifically kill cancer cells without significant effect on normal cells (where only non-essential activity of *B* is suppressed, and *A* is intact). Gene *A* can be a tumor suppressor gene: for example, a genome caretaker. Using synthetic lethality approach, the mutated and non-functional tumor suppressor genes can be targeted indirectly through their genetic interactions.

One of the drugs currently passing the clinical trials and based on this principle is a family of so called PARP-inhibitors (Helleday, 2011). PARP1 protein is known to be involved in Base-excision DNA repair pathway (among many other functions), which is capable of fixing the single-strand breaks of DNA. When PARP1 function is suppressed, the single-strand breaks can lead to double-strand breaks of DNA, which are lethal for a cell and are usually repaired by BRCA-dependent homologous recombination (HR) DNA repair pathway. If a tumor contains mutations invalidating both alleles of BRCA, this, together with knocking-down PARP, can lead to cell death due to unrepaired double-strand DNA breaks. This explanation is not complete because double-strand DNA breaks can be also repaired by an alternative error-prone mechanism, called Non-Homologous End Joining (NHEJ). In reality, the structure of DNA repair network is even more complex. To have an idea of the realistic complexity of DNA repair network, as well as of the crosstalk between different DNA repair pathways, and their connection to cell-cycle through cell-cycle checkpoints, we have constructed the comprehensive map of DNA repair, available as a part of Atlas of Cancer Signaling Networks (ACSN, https://acsn.curie.fr ).

PARP inhibition is only a pilot example of using synthetic lethality as a therapeutic principle in cancer. We believe that this principle should have much wider application. One can imagine that in the future, a target tumor tissue to be treated will be profiled for mutations in all genes. Following, this profiling, the therapeutic combinations of drugs will be composed of such inhibitors of protein function that their targets will form synthetic-lethal combinations with one or several detected mutations. For implementations of such cancer treatment program, one needs to have a comprehensive list of synthetic-lethal interactions. It is in principle possible to obtain such a list from large-scale screenings for viability of double-mutant cell colonies. However, until so far, the genome-wide screening approach has been limited to model organisms (Costanzo et al, 2010). In mammalian cells, the screenings are performed for testing synthetic-lethal interactions limited to some frequently mutated genes, using siRNA-based knock-downs.

However, the possibility of the experimental approach is limited in many aspects. One of them is that in higher organisms, due to higher redundancy, the synthetic-lethal combinations can be composed not only from pairs but from gene triplets, or gene 4-tuples, and so on. Testing all these combinations will not be feasible in the nearest future. Therefore, we need theoretical tools allowing highlighting those combinations of genes that show lethal effect. This problem can be approached in different ways. We have decided to predict probable synthetic-lethal combinations from comprehensive reconstructions of reaction networks involved in cancer. Until so far, we obtained several theoretical results: 1) we systematically described and have catalogued known molecular mechanisms of synthetic lethality (Kuperstein et al, unpublished); 2) we described a new type of molecular mechanism leading to synthetic lethality, called “a kinetic trap” (Zinovyev et al, 2013B); 3) we developed an algorithm and software for predicting synthetic-lethal combinations, resulting from the simplest mechanism of path redundancy (Vera-Licona et al, 2013); 4) we decided to focus on predicting synthetic lethality combinations in mammalian DNA repair network and develop theoretical tools for this. Below I will briefly focus on my investigations of synthetic lethality in DNA repair.

Focusing on DNA repair for studying synthetic lethality relevant for cancer treatment is justified for many reasons. Germline and somatic mutations in DNA repair genes (such as ATM, MSH2, BRCA1) are frequently found in many types of cancers. Many of these genes are “genome caretakers”, participating in assuring genome stability which is defined as major “enabling characteristics” of cancer (Hanahan and Weinberg, 2011).

Affecting DNA repair efficiency by loss-of-function mutations is a double-edge sword for cancer. From one hand, increasing genomic instability leads to increasing genomic diversity and, hence, to higher evolvability of cancer cells. From another hand, defects in DNA repair can be lethal for the cell, either because of the defects in DNA incompatible with cell functioning, or because of triggering apoptosis by cell-cycle checkpoint mechanisms.

One of the most dangerous defects in DNA is a double-strand break. Left unrepaired, it is usually lethal for the cell. One of the DNA repair pathways capable to repair the double-strand breaks is already above-mentioned Homologous Recombination (HR) pathway. I was intrigued by the fact that in yeast some double knockouts of the genes involved in HR are lethal (for example, knocking out Srs2 and Rad54, (Palladino and Klein, 1992)). It is counter-intuitive for two reasons. First of all, HR pathway is not *an essential pathway* in yeast and can be compensated by other pathways.  Second, synthetic-lethality between genes *in the same pathway* does not follow the classical paradigm of the between-redundant-pathway model. Studying this question, it was noticed that some of the transitions between repairing DNA states *are reversible and regulated*: for example, Srs2 regulates one of the backward transitions. This led to the idea of explaining this synthetic lethality by the “kinetic trap” model (Figure 10).  We analyzed this situation formally, exploiting mathematical modeling, and clarified the quantitative aspects of this mechanism (Zinovyev et al, 2013B). Moreover, we hypothesized that the kinetic trap mechanism can be responsible for appearance of synthetic lethality in other molecular processes, not necessarily related to DNA repair. For example, kinetic trapping can make an essential cascade of reversible post-translational protein modifications being trapped in a state not allowing the signal to propagate.

In a complementary approach, we used the comprehensive reconstruction of the DNA repair network (available at https://acsn.curie.fr/navicell/maps/dnarepair/master/index.html) to analyze the structure of the graph describing all possible transitions between the states of repairing DNA, and their regulators. We focused at that part of the graph which describes possible appearance of double-strand breaks, either from unrepaired single-strand breaks or directly as an action of certain drugs. Our analysis showed very few possibilities to disrupt this part of DNA network by pairwise gene knock-outs, while many possibilities to do it by knocking out triplets of genes (Kuperstein et al, unpublished). In particular, we concluded that deletion of BRCA1 and PARP protein functions can be synthetically lethal only in some specific cellular contexts, where the function of certain other DNA repair proteins is impaired. We suggested that this analysis can explain part of the controversial experimental and clinical observations on the sensitivity of cells to PARP inhibitors (Helleday, 2011).

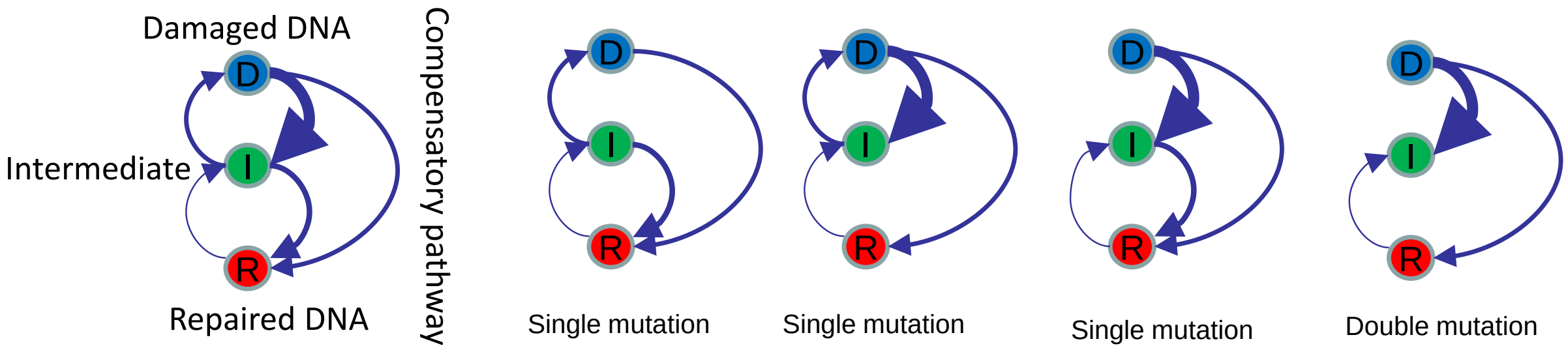

Figure 10. The simplest model of kinetic trap (for example, in HR pathway), leading to synthetic lethality between two enzymes participating in the same non-essential pathway. The edge thickness denotes the relative speeds between different states of DNA (damaged, intermediate, repaired). It is shown that a single mutation is not able to make the mechanism non-functional, due to the presence of a compensatory mechanism and a backward reaction from the intermediate state. However, there exists a combination of two mutations in the pathway capturing it in a non-functional and, may be, toxic state.

# *5. CONCLUSION AND PERSPECTIVES*

The conclusions from my experience of working with the problems where the biological complexity is an essential feature are rather optimistic: the complexity of biological systems, as they are represented by formal mathematical models, can be fought with. Their complexity is frequently only seaming, representing rather complicatedness of the system, its ability to function in many different situations. However, each particular system's response to a perturbation can be simple and comprehensible. Thus, the complexity can be reduced by dissection or simplification, provided appropriate experimental and computational methods.

This conclusion is a pleasant surprise rather than an expectation: one could have assumed that the natural selection has tendency to "complexify" biological systems, pushing them to the "wild" complexity situation. My conclusion can be misleading as well, and related only to our way of representing the biological reality by experimental data or mathematical models. It might happen that these models or data are not adequate to grasp the essence and the essential complexity of a living cell. However, I do not think so.

Of course, we are very far from feeling comfortable being faced with the biological complexity. We lack approaches allowing working and simulating large and incompletely characterized biochemical networks. We do not know how to systematically integrate the existing data in the most useful way. Most importantly, we do not know how to use the available data in the most efficient way using our knowledge about cellular molecular mechanisms, how to efficiently produce new knowledge on the functioning of molecular mechanisms from the large-scale genomic-wide data. Lack of good and objectively validated approaches in these fields is reflected in difficulties with realistic and large-scale modeling of cells, tissues, organs, organisms.

Nevertheless, the progress in quantifying, understanding and reducing complexity of biological systems is real and inevitable. Recent breakthroughs in multi-level profiling of molecular composition of a cell, whole-cell computational models, and realistic multi-scale models describing both biochemical and biophysical properties of cell populations make this direction of scientific research exciting and full of promises.

# *REFERENCES* [1]

[1] Sorted by the first author name and the year of the publication